# THE ROTATING AND ACCELERATING UNIVERSE

**Evangelos Chaliasos**

365 Thebes Street

GR-12241 Aegaleo

Athens, Greece

*Abstract*

An attempt is made to explain the spiral structure of spiral galaxies through a possible rotation of the Universe. To this end, we write down a possible form of the metric, and we calculate the necessary quantities ($R_{ik}$, $T_{ik}$, …) in order to form the Einstein equations. We find the two Einstein equations pertaining to the Robertson-Walker metric and no rotation at all in this way, if we assume $p+\varepsilon$ different from 0.

There are introduced then two suitable rotational motions in the universe, and we try to generalize again the Robertson-Walker metric in this way. The result is again null, since it is found that the corresponding angular velocities must vanish, for not vanishing $p+\varepsilon$.

A third attempt is finally done, without restricting ourselves to generalize any existing cosmological model. We introduce again two suitable rotational motions, and we form the appropriate Einstein equations. After solving them, we find that the two "rotations", corresponding to the usual rotations on a torus, dictate again the equation of state $p+\varepsilon=0$. This solution is a steady state one, and it describes naturally the observationally discovered acceleration of the Universe (1998), *without a cosmological constant and the ambiguous "dark energy".*

## 1. Introduction

The spiral structure of spiral galaxies constitutes the most wonderful phenomenon of the night sky perhaps. A lot of theories have been developed in order to explain it (see e.g. Binnay & Tremaine, 1987). The most prominent of these theories is the density wave theory of Lin & Shu. These scientists made the hypothesis that "spiral structure consists of a quasi-steady density wave, that is, a density wave that is maintained in a steady state over many rotation periods". They "did not at first address the question of the origin of the spiral structure". Besides, "it has proved to be frustratingly difficult to determine unambiguously the basic parameters of the models" produced after adoption of the Lin-Shu hypothesis of density waves. In fact "ironically, the Lin-Shu hypothesis, … , may still prove to be largely irrelevant to spiral structure" (!) (Binnay & Tremaine, 1987).

Now, the thing which comes to mind when we observe a picture of a spiral galaxy (Fig.1) is unambiguously the picture of a cyclone (Fig.2). In fact, the similarity of the two pictures is striking. But we know that cyclones are a result of the Coriolis forces, caused by the rotation of Earth. It is natural then for someone to attribute the spiral structure of spiral galaxies to similar Coriolis forces, resulting from a kind of rotation of the Universe as a whole. This is the motivation underlying the present paper.

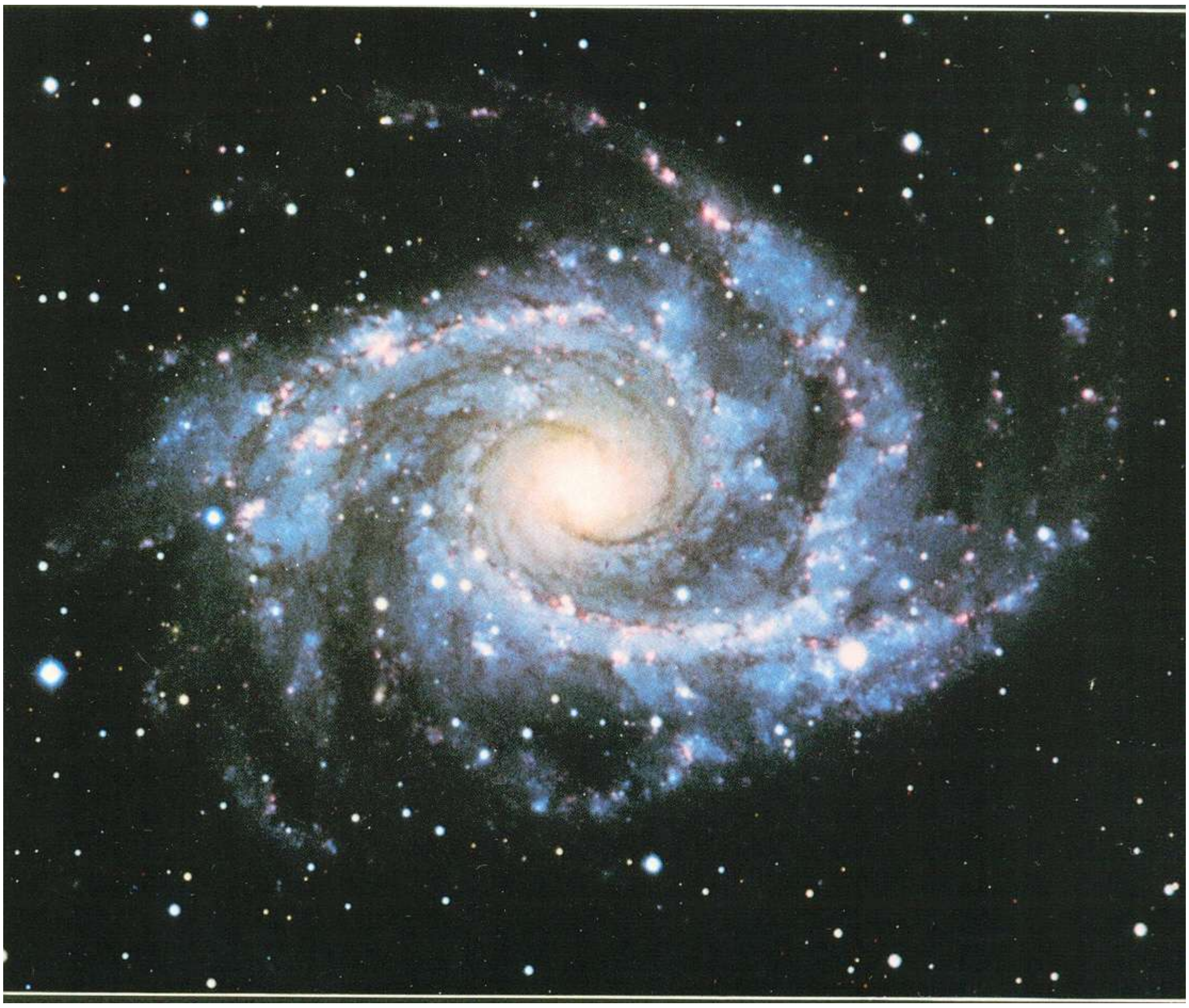

Fig1. A spiral galaxy

The most well-known exact solution of the Einstein equations representing a rotating model universe is Goedel´s universe (see, e.g., Hawking & Ellis, 1973). But it is not acceptable for it admits of closed time-like curves, causality in this way being violated.

Much of work around the Goedel universe has been done by M.J. Reboucas and his collaborators about 1980 (Novello & Reboucas, 1978a; Reboucas, 1978b; Novello & Reboucas, 1979; Reboucas & de Lima, 1981; Reboucas & Tiomno, 1983). This kind of work was continued by L.K. Patel and his co-workers (Vaidya & Patel, 1984; Koppar & Patel, 1988a; Koppar & Patel, 1988b).

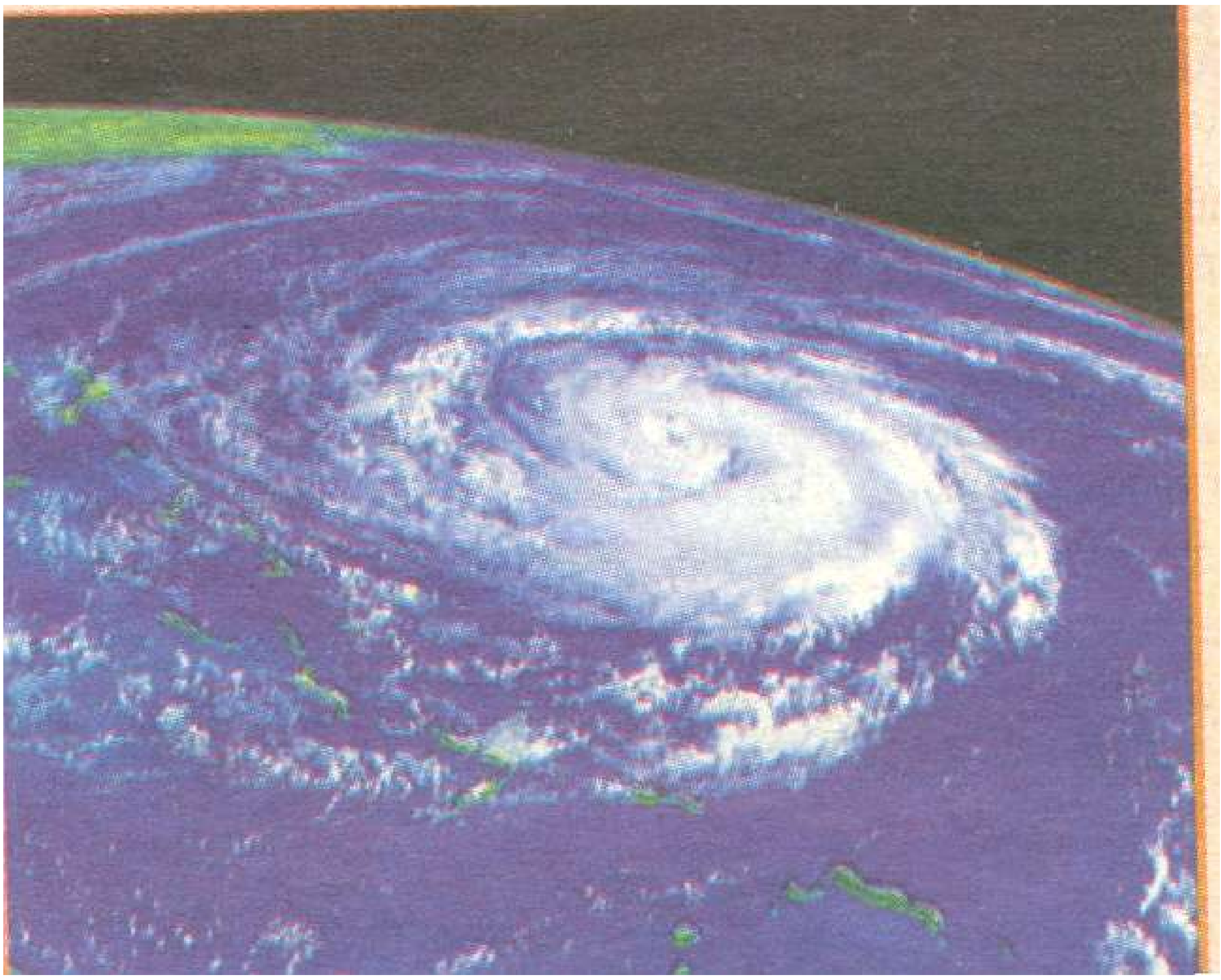

Fig.2. A cyclone

It has to be noticed that a rotating universe stops of being isotropic, remaining perhaps homogeneous. Thus a study of homogeneous spaces and the various Bianchi types arising seems to be necessary (Landau & Lifshitz, 1975; Ryan & Shepley, 1975). This study of exact solutions of Einstein´s equations (Kramer *et al,* 1980) is formally done by the method of Lie groups (Chevalley, 1970). In this sense work has been done by Rosquist, 1983.

For cosmological rotation in Brans-Dicke Theory see Usham & Tarachand Singh, 1992. For rotating and expanding cosmology in ECSK-Theory see Korotkii & Obukhov, 1992.

My own initial work, e.g. the first attempt (described in section 2) did not use the advanced tools described above, but only plain classical General Relativity, as it will be seen in what follows, and it failed. In section 3, I did a second attempt, but unfortunately unsuccessful again. After that, in the remainder of my paper, I used too the techniques of section 3, and I took again p+ε=0, in order to save rotation.

In what follows, the sections (or the subsections) marked with the symbol * (an asterisk) may be omitted in a first reading, serving merely as appendices.

## 2. The simply rotating universe *

See Appendix I.

## 3. The doubly rotating universe *

See AppendixII.

## 4. The doubly rotating Universe continued

*i) The form of the metric*

Instead of taking the Robertson-Walker line element as the starting point, I now take as the starting point a more general line element. Namely, I consider again our three-dimensional space as a three-dimensional hypersurface embedded in a fictitious four-dimensional Euclidean space with coordinates $x_1$, $x_2$, $x_3$, $x_4$, and I assume that our hypersurface has again <u>two</u> rotational (cylindrical) symmetries: one on the plane ($x_1$, $x_2$), and the other on the plane ($x_3$, $x_4$), so that equations (9) are satisfied again. Thus, I will have again

$$x_1^2 + x_2^2 = a_1^2 \quad \text{(a)} \quad \& \quad x_3^2 + x_4^2 = a_2^2 \quad \text{(b)} \qquad (29)$$

Though, I will not take now, as in eqn. (10),

$$(x_1^{\ 2}+x_2^{\ 2})+(x_3^{\ 2}+x_4^{\ 2})=a^2 \tag{30}$$

(which would lead to the Robertson-Walker case again) but instead I will take $x_1^2+x_2^2$ and $x_3^2+x_4^2$ to satisfy a more general equation of the form

$$F(x_1^{\ 2}+x_2^{\ 2},x_3^{\ 2}+x_4^{\ 2})=C, \tag{31}$$

where C is a constant. Then the family of curves (31) can be written in parametric form as

$$x_1^{\ 2}+x_2^{\ 2}=f_1(\lambda,C) \quad \text{(a)} \quad \& \quad x_3^{\ 2}+x_4^{\ 2}=f_2(\lambda,C) \quad \text{(b)} \tag{32}$$

with λ the parameter and $f_1$ & $f_2$ two suitable functions. Thus, I will have instead of eqns. (11)

$$a_1^{\ 2}=f_1(\lambda,C) \quad \text{(a)} \quad \& \quad a_2^{\ 2}=f_2(\lambda,C) \quad \text{(b)} \tag{32′}$$

so that the family (31) will become

$$F\left(a_1^{\ 2}(\lambda,C),a_2^{\ 2}(\lambda,C)\right)=C. \tag{33}$$

The new coordinates of the 4-space will now be C, λ, $\varphi_1$, $\varphi_2$, with

$$\left.\begin{aligned} x_1 &= a_1(\lambda,C)\cos\phi_1 \\ x_2 &= a_1(\lambda,C)\sin\phi_1 \\ x_3 &= a_2(\lambda,C)\cos\phi_2 \\ x_4 &= a_2(\lambda,C)\sin\phi_2 \end{aligned}\right\} \tag{34}$$

Differentiating the coordinate transformation (34), and adding up the squares of the differentials of the old coordinates, we find for the line element of 4-space (14)

$$dL^2=\left(\frac{\partial a_1}{\partial\lambda}d\lambda+\frac{\partial a_1}{\partial C}dC\right)^2+a_1^{\ 2}d\phi_1^{\ 2}+\left(\frac{\partial a_2}{\partial\lambda}d\lambda+\frac{\partial a_2}{\partial C}dC\right)^2+a_2^{\ 2}d\phi_2^{\ 2}. \tag{35}$$

The line element dl of our hypersurface will then be given by eqn. (35) for dC=0. Thus

$$dl^2=\left[\left(\frac{\partial a_1}{\partial\lambda}\right)^2+\left(\frac{\partial a_2}{\partial\lambda}\right)^2\right]d\lambda^2+a_1^{\ 2}d\phi_1^{\ 2}+a_2^{\ 2}d\phi_2^{\ 2}. \tag{36}$$

The line element of the 4-dimensional space-time, if we choose it to be *synchronous,* will be

$$ds^2 = c^2dt^2 - dl^2, \qquad (37)$$

with $dl^2$ given by (36).

In order now to account for the double rotation of the Universe, it is enough to let

$$\left.\begin{array}{ll} d\phi_1 \to d\phi_1 - \omega_1 dt & \text{(a)} \\ d\phi_2 \to d\phi_2 - \omega_2 dt & \text{(b)} \end{array}\right\} \qquad (38)$$

as we did in the previous section and for the same reason. I note that, in our case, I can set C=C(t). Then $a_1=a_1(\lambda, t)$ and $a_2=a_2(\lambda, t)$. I similarly let $\omega_1=\omega_1(\lambda, t)$ and $\omega_2=\omega_2(\lambda, t)$. Then I will finally obtain for the form of the metric for this doubly rotating Universe

$$ds^2 = c^2dt^2 - \left[\left(\frac{\partial a_1}{\partial \lambda}\right)^2 + \left(\frac{\partial a_2}{\partial \lambda}\right)^2\right]d\lambda^2 - a_1^{\,2}(d\phi_1 - \omega_1 dt)^2 - a_2^{\,2}(d\phi_2 - \omega_2 dt)^2, \qquad (39)$$

with $a_1$ & $a_2$ and $\omega_1$ & $\omega_2$ being functions of $\lambda$ and t, to be determined via the Einstein equations.

*ii) The components of the metric tensor*

If I expand the metric (39), I obtain

$$ds^2 = \left(1 - a_1^{\,2}\frac{\omega_1^{\,2}}{c^2} - a_2^{\,2}\frac{\omega_2^{\,2}}{c^2}\right)c^2dt^2 + 2a_1^{\,2}\frac{\omega_1}{c}cdtd\phi_1 + 2a_2^{\,2}\frac{\omega_2}{c}cdtd\phi_2 -$$

$$- \left[\left(\frac{\partial a_1}{\partial \lambda}\right)^2 + \left(\frac{\partial a_2}{\partial \lambda}\right)^2\right]d\lambda^2 - a_1^{\,2}d\phi_1^{\,2} - a_2^{\,2}d\phi_2^{\,2}. \qquad (40)$$

Thus, the non-vanishing covariant components $g_{ik}$ of the metric tensor are

$$g_{00} = 1 - a_1^{\,2}\frac{\omega_1^{\,2}}{c^2} - a_2^{\,2}\frac{\omega_2^{\,2}}{c^2}$$

$$g_{11} = -\left[\left(\frac{\partial a_1}{\partial \lambda}\right)^2 + \left(\frac{\partial a_2}{\partial \lambda}\right)^2\right]$$

$$g_{22} = -a_1^{\,2}$$

$$g_{33} = -a_2^{\,2}$$

$$g_{02} = g_{20} = a_1^{\,2}\frac{\omega_1}{c}$$

$$g_{03} = g_{30} = a_2^{\,2}\frac{\omega_2}{c}. \qquad (41)$$

In order to find the contravariant components gik we have to find the determinant g and the relevant determinants Gik. The determinant g is found to be

$$g = -a_1^2 a_2^2 \left[ \left( \frac{\partial a_1}{\partial \lambda} \right)^2 + \left( \frac{\partial a_2}{\partial \lambda} \right)^2 \right] \tag{42}$$

The non vanishing relative determinants Gik are found to

$$G^{00} = -a_1^2 a_2^2 \left[ \left( \frac{\partial a_1}{\partial \lambda} \right)^2 + \left( \frac{\partial a_2}{\partial \lambda} \right)^2 \right]$$

$$G^{02} = G^{20} = -a_1^2 a_2^2 \frac{\omega_1}{c} \left[ \left( \frac{\partial a_1}{\partial \lambda} \right)^2 + \left( \frac{\partial a_2}{\partial \lambda} \right)^2 \right]$$

$$G^{03} = G^{30} = -a_1^2 a_2^2 \frac{\omega_2}{c} \left[ \left( \frac{\partial a_1}{\partial \lambda} \right)^2 + \left( \frac{\partial a_2}{\partial \lambda} \right)^2 \right]$$

$$G^{11} = a_1^2 a_2^2$$

$$G^{22} = \left( 1 - a_1^2 \frac{\omega_1^2}{c^2} \right) a_2^2 \left[ \left( \frac{\partial a_1}{\partial \lambda} \right)^2 + \left( \frac{\partial a_2}{\partial \lambda} \right)^2 \right]$$

$$G^{23} = G^{32} = -a_1^2 a_2^2 \frac{\omega_1 \omega_2}{c^2} \left[ \left( \frac{\partial a_1}{\partial \lambda} \right)^2 + \left( \frac{\partial a_2}{\partial \lambda} \right)^2 \right]$$

$$G^{33} = \left( 1 - a_2^2 \frac{\omega_2^2}{c^2} \right) a_1^2 \left[ \left( \frac{\partial a_1}{\partial \lambda} \right)^2 + \left( \frac{\partial a_2}{\partial \lambda} \right)^2 \right] \tag{43}$$

Thus, the non-vanishing contravariant components $g^{ik}$ of the metric tensor, given by

$$g^{ik} = G^{ik}/g, \tag{44}$$

are found to be

$$g^{00}=1$$
$$g^{02}=g^{20}=\omega_1/c$$
$$g^{03}=g^{30}=\omega_2/c$$
$$g^{11}=-\left[\left(\frac{\partial a_1}{\partial\lambda}\right)^2+\left(\frac{\partial a_2}{\partial\lambda}\right)^2\right]^{-1}$$
$$g^{22}=-\frac{1}{a_1^{\;2}}+\frac{\omega_1^{\;2}}{c^2}$$
$$g^{23}=g^{32}=\omega_1\omega_2/c^2$$
$$g^{33}=-\frac{1}{a_2^{\;2}}+\frac{\omega_2^{\;2}}{c^2}. \tag{45}$$

*iii)The tetrad and the associated one-forms*

I write the metric (39) in the form (Landau & Lifshitz, 1975; Chandrasekhar,1992)

$$ds^2=\eta_{ab}\left(e^{(a)}{}_i dx^i\right)\left(e^{(b)}{}_k dx^k\right) \tag{46}$$

with $\eta_{ab}$ a suitable constant matrix. The associated one-forms (Chandrasekhar, 1992; Lovelock & Rund, 1975) will be

$$\omega^a=e^{(a)}{}_i dx^i. \tag{47}$$

I take as the most natural choice (cf. eqn. (39))

$$\omega^0=cdt=e^{(0)}{}_i dx^i$$
$$\omega^1=\left[\left(\frac{\partial a_1}{\partial\lambda}\right)^2+\left(\frac{\partial a_2}{\partial\lambda}\right)^2\right]^{1/2}d\lambda=e^{(1)}{}_i dx^i$$
$$\omega^2=a_1\left(d\phi_1-\omega_1 dt\right)=e^{(2)}{}_i dx^i$$
$$\omega^3=a_2\left(d\phi_2-\omega_2 dt\right)=e^{(3)}{}_i dx^i \tag{48}$$

The covariant components of the "contravariant" tetrad basis vectors are then

$$e^{(0)}{}_i=(1,0,0,0)$$
$$e^{(1)}{}_i=\left(0,\left[\left(\frac{\partial a_1}{\partial\lambda}\right)^2+\left(\frac{\partial a_2}{\partial\lambda}\right)^2\right]^{1/2},0,0\right)$$
$$e^{(2)}{}_i=\left(-a_1\frac{\omega_1}{c},0,a_1,0\right)$$
$$e^{(3)}{}_i=\left(-a_2\frac{\omega_2}{c},0,0,a_2\right). \tag{49}$$

By the use of the contravariant components of the metric tensor (45), we find from (49) for the contravariant components of the “contravariant” tetrad basis vectors

$$
\begin{aligned}
e^{(0)i} &= \left(1, 0, \omega_1/c, \omega_2/c\right) \\
e^{(1)i} &= \left(0, -\left[\left(\frac{\partial a_1}{\partial \lambda}\right)^2 + \left(\frac{\partial a_2}{\partial \lambda}\right)^2\right]^{-1/2}, 0, 0\right) \\
e^{(2)i} &= \left(0, 0, -1/a_1, 0\right) \\
e^{(3)i} &= \left(0, 0, 0, -1/a_2\right)
\end{aligned}
\qquad (50)
$$

We obtain that

$$
\left.
\begin{aligned}
e^{(0)}{}_i e^{(0)i} &= +1 \\
e^{(1)}{}_i e^{(1)i} &= -1 \\
e^{(2)}{}_i e^{(2)i} &= -1 \\
e^{(3)}{}_i e^{(3)i} &= -1
\end{aligned}
\right\}
\qquad (51)
$$

and

$$e^{(a)}{}_i e^{(b)i} = 0, \quad \text{for} \quad a \neq b. \qquad (52)$$

Thus, from

$$e^{(a)}{}_i e^{(b)i} = \eta^{ab}, \qquad (53)$$

we obtain

$$\eta^{ab} = diag(1, -1, -1, -1), \qquad (54)$$

that is Minkowskian.

I will proceed now to the determination of the “covariant” tetrad basis vectors, and their contravariant tensor components $e_{(b)}{}^i$ at first. They are given by

$$e^{(a)}{}_{i}e_{(b)}{}^{i} = \delta^{a}{}_{b}, \tag{55}$$

where $\delta^a{}_b$ are the Kronecker δ-symbols. Thus, the matrix $e_{(b)}{}^i$ is the reciprocal (inverse) of the matrix

$$e^{(a)}{}_{i} = \begin{pmatrix} 1 & 0 & 0 & 0 \\ 0 & \left[\left(\frac{\partial a_1}{\partial\lambda}\right)^2 + \left(\frac{\partial a_2}{\partial\lambda}\right)^2\right]^{1/2} & 0 & 0 \\ -a_1\frac{\omega_1}{c} & 0 & a_1 & 0 \\ -a_2\frac{\omega_2}{c} & 0 & 0 & a_2 \end{pmatrix} \tag{56}$$

I find for its determinant

$$\Delta = a_1 a_2 \left[\left(\frac{\partial a_1}{\partial\lambda}\right)^2 + \left(\frac{\partial a_2}{\partial\lambda}\right)^2\right]^{1/2}. \tag{57}$$

The relevant determinants $E_{(b)}{}^i$ are also needed. I find that the non-vanishing of them are given by

$$\begin{aligned}
E_{(0)}{}^{0} &= a_1 a_2 \left[\left(\frac{\partial a_1}{\partial\lambda}\right)^2 + \left(\frac{\partial a_2}{\partial\lambda}\right)^2\right]^{1/2} \\
E_{(0)}{}^{2} &= a_1 a_2 \frac{\omega_1}{c} \left[\left(\frac{\partial a_1}{\partial\lambda}\right)^2 + \left(\frac{\partial a_2}{\partial\lambda}\right)^2\right]^{1/2} \\
E_{(0)}{}^{3} &= a_1 a_2 \frac{\omega_2}{c} \left[\left(\frac{\partial a_1}{\partial\lambda}\right)^2 + \left(\frac{\partial a_2}{\partial\lambda}\right)^2\right]^{1/2} \\
E_{(1)}{}^{1} &= a_1 a_2 \\
E_{(2)}{}^{2} &= a_2 \left[\left(\frac{\partial a_1}{\partial\lambda}\right)^2 + \left(\frac{\partial a_2}{\partial\lambda}\right)^2\right]^{1/2} \\
E_{(3)}{}^{3} &= a_1 \left[\left(\frac{\partial a_1}{\partial\lambda}\right)^2 + \left(\frac{\partial a_2}{\partial\lambda}\right)^2\right]^{1/2}
\end{aligned} \tag{58}$$

The non-vanishing contravariant tensor components of the "covariant" tetrad basis vectors, given by

$$e_{(b)}{}^{i} = E_{(b)}{}^{i} / \Delta, \tag{59}$$

are found to be

$$
\begin{aligned}
e_{(0)}{}^{i} &= \left(1,0,\omega_1/c,\omega_2/c\right)\\
e_{(1)}{}^{i} &= \left(0,\left[\left(\frac{\partial a_1}{\partial\lambda}\right)^2+\left(\frac{\partial a_2}{\partial\lambda}\right)^2\right]^{-1/2},0,0\right)\\
e_{(2)}{}^{i} &= \left(0,0,1/a_1,0\right)\\
e_{(3)}{}^{i} &= \left(0,0,0,1/a_2\right),
\end{aligned}
\tag{60}
$$

having as a matrix

$$
\begin{pmatrix}
1 & 0 & 0 & 0\\
0 & \left[\left(\frac{\partial a_1}{\partial\lambda}\right)^2+\left(\frac{\partial a_2}{\partial\lambda}\right)^2\right]^{-1/2} & 0 & 0\\
\omega_1/c & 0 & 1/a_1 & 0\\
\omega_2/c & 0 & 0 & 1/a_2
\end{pmatrix}
\tag{61}
$$

From the matrices (56) and (61) we confirm that

$$
e^{(a)}{}_{i}e_{(a)}{}^{k} = \delta_i^{\,k}, \tag{62}
$$

as it should be.

With the aid of the covariant components (41) of the metric tensor, I finally find for the covariant tensor components of the "covariant" tetrad basis vectors, that they are given by

$$
\begin{aligned}
e_{(0)i} &= (1,0,0,0)\\
e_{(1)i} &= (0,-\left[\left(\frac{\partial a_1}{\partial\lambda}\right)^2+\left(\frac{\partial a_2}{\partial\lambda}\right)^2\right]^{1/2},0,0)\\
e_{(2)i} &= (a_1\omega_1/c,0,-a_1,0)\\
e_{(3)i} &= (a_2\omega_2/c,0,0,-a_2).
\end{aligned}
\tag{63}
$$

I confirm that

$$
\left.
\begin{aligned}
e_{(0)}{}^{i}e_{(0)i} &= +1\\
e_{(1)}{}^{i}e_{(1)i} &= -1\\
e_{(2)}{}^{i}e_{(2)i} &= -1\\
e_{(3)}{}^{i}e_{(3)i} &= -1
\end{aligned}
\right\}
\tag{64}
$$

and

$$e_{(a)}{}^{i} e_{(b)i} = 0, \quad \text{for} \quad a \neq b, \tag{65}$$

that is

$$e_{(a)}{}^{i} e_{(b)i} = \eta_{ab}, \tag{66}$$

with

$$\eta_{ab} = diag(1,-1,-1,-1), \tag{67}$$

as expected (cf. (54)).

*iv) The curvature tensor via Cartan΄s equations of structure **

See Appendix III.

*v) The curvature tensor via the λ-symbols of tetrad formalism **

See AppendixIV.

*vi) The energy-momentum tensor and Einstein΄s equations**

See Appendix V.

*vii) Solution of the Einstein equations and the metric*

α΄) The {01}, {12}, {13} Einstein equations

I set

$$a \equiv [\ ]^{1/2} = \left[ \left( \frac{\partial a_1}{\partial \lambda} \right)^2 + \left( \frac{\partial a_2}{\partial \lambda} \right)^2 \right]^{1/2}. \tag{169}$$

I also set

$$a_1(\lambda,t) \equiv b_1(t)c_1(\lambda)$$
$$a_2(\lambda,t) \equiv b_2(t)c_2(\lambda), \qquad (170)$$

that is I assume that $a_1$ & $a_2$ are separable in λ & t. Then I have

$$\frac{\partial a_1}{\partial\lambda} = b_1 c_1' \quad \& \quad \frac{\partial a_2}{\partial\lambda} = b_2 c_2', \qquad (171)$$

so that

$$a = \left[b_1^2 c_1'^2 + b_2^2 c_2'^2\right]^{1/2}, \quad \text{and} \qquad (172)$$

$$\frac{\partial}{c\partial t}\ln a = \frac{b_1\dot{b}_1 c_1'^2 + b_2\dot{b}_2 c_2'^2}{c\left[b_1^2 c_1'^2 + b_2^2 c_2'^2\right]}. \qquad (173)$$

The dots (primes) will mean differentiation with respect to t (λ). I also have

$$\frac{\partial}{\partial\lambda}\ln a_1 = \frac{\partial}{\partial\lambda}\ln c_1 \quad \text{or} \quad \frac{c_1'}{c_1}$$
$$\frac{\partial}{\partial\lambda}\ln a_2 = \frac{\partial}{\partial\lambda}\ln c_2 \quad \text{or} \quad \frac{c_2'}{c_2}, \qquad (174)$$
$$\frac{\partial}{c\partial t}\ln a_1 = \frac{\partial}{c\partial t}\ln b_1 \quad \text{or} \quad \frac{\dot{b}_1}{cb_1}$$
$$\frac{\partial}{c\partial t}\ln a_2 = \frac{\partial}{c\partial t}\ln b_2 \quad \text{or} \quad \frac{\dot{b}_2}{cb_2}, \quad \text{and} \qquad (175)$$
$$\frac{\partial^2}{\partial\lambda c\partial t}\ln a_1 = 0$$
$$\frac{\partial^2}{\partial\lambda c\partial t}\ln a_2 = 0. \qquad (176)$$

Now, the {01} Einstein equation can be written

$$\left\{\frac{b_1\dot{b}_1 c_1'^2 + b_2\dot{b}_2 c_2'^2}{b_1^2 c_1'^2 + b_2^2 c_2'^2} - \frac{\dot{b}_1}{b_1}\right\}\frac{c_1'}{c_1} + \left\{\frac{b_1\dot{b}_1 c_1'^2 + b_2\dot{b}_2 c_2'^2}{b_1^2 c_1'^2 + b_2^2 c_2'^2} - \frac{\dot{b}_2}{b_2}\right\}\frac{c_2'}{c_2} = 0. \qquad (177)$$

After some manipulations, I find

$$\frac{c_1' c_2' (b_1\dot{b}_2 - b_2\dot{b}_1)(b_2^2 c_2 c_2' - b_1^2 c_1 c_1')}{b_1 b_2 c_1 c_2 [b_1^2 c_1'^2 + b_2^2 c_2'^2]} = 0, \qquad (178)$$

and thus I have

$$\left.\begin{array}{ll} b_1\dot{b}_2 = b_2\dot{b}_1, & \text{(a)} \quad \text{or} \\ b_2^2 c_2 c_2' = b_1^2 c_1 c_1' & \text{(b)} \end{array}\right\} \qquad (179)$$

From eqn. (a), upon integration, I get

$$\Theta b_1(t) = b_2(t), \qquad (180)$$

with Θ a constant. From eqn. (b) I get

$$\left.\begin{aligned} b_1^2(t) &= \Theta^{-2} b_2^2(t), \quad \text{(a)} \quad \& \\ c_2 dc_2 &= \Theta^{-2} c_1 dc_1. \quad \text{(b)} \end{aligned}\right\} \tag{181}$$

Thus, I always have

$$b_1(t) = \Theta^{-1} b_2(t), \tag{182}$$

which is necessary and sufficient for the validity of eqn. (177). Then

$$\frac{a_1}{a_2} = \frac{b_1}{b_2}\frac{c_1}{c_2} = \Theta^{-1}\frac{c_1(\lambda)}{c_2(\lambda)}. \tag{183}$$

Therefore, for constant λ, it follows that $a_1/a_2$ = constant. That is I can take as λ the angle between the vector ($a_1$, $a_2$) and the $a_1$-axis in the ($a_1$, $a_2$) plane. Then, because

$$a_2 / a_1 = \tan\lambda, \tag{184}$$

it follows that

$$\frac{1}{\tan\lambda} = \Theta^{-1}\frac{c_1(\lambda)}{c_2(\lambda)}, \tag{185}$$

that is

$$\frac{c_2(\lambda)}{c_1(\lambda)} = \frac{\tan\lambda}{\Theta}. \tag{186}$$

It is easy to see that the curves of the family F = C (each of them corresponding to differrent t) are homeothetous to one another in the ($a_1$, $a_2$) plane.

For constant t (same C), we have

$$\left.\begin{aligned} a_1 &= b_1 c_1 \equiv b(\lambda)c(\lambda), \quad \text{(a)} \quad \text{and} \\ a_2 &= b_2 c_2 \equiv \Theta b(t)\frac{\tan\lambda}{\Theta}c(\lambda), \quad \text{(b)} \end{aligned}\right\} \tag{187}$$

that is

$$\left.\begin{aligned} a_1 &= b(t)c(\lambda), \quad \text{(a)} \quad \text{and} \\ a_2 &= b(t)c(\lambda)\tan\lambda. \quad \text{(b)} \end{aligned}\right\} \tag{188}$$

Set

$$\left.\begin{aligned} a_1 &= \tilde{a}\cos\lambda \quad \text{(a)} \\ a_2 &= \tilde{a}\sin\lambda, \quad \text{(b)} \end{aligned}\right\} \tag{189}$$

where a~ is the magnitude of the vector ($a_1$, $a_2$) on the ($a_1$, $a_2$) plane. We thus have

$$\tilde{a} = b\frac{c}{\cos\lambda} \equiv \tilde{b}(t)\tilde{c}(\lambda). \tag{190}$$

Set

$$\left.\begin{array}{ll} \tilde{b}(t) \equiv u(t) \quad (= b), & \text{(a)} \quad \text{and} \\ \tilde{c}(t) \equiv v(\lambda) \quad \left(= \dfrac{\text{c}}{\cos\lambda}\right) & \text{(b)} \end{array}\right\} \tag{191}$$

Then eqns. (189) can be written

$$\left.\begin{array}{ll} a_1(\lambda,t) = u(t)v(\lambda)\cos\lambda & \text{(a)} \\ a_2(\lambda,t) = u(t)v(\lambda)\sin\lambda\,. & \text{(b)} \end{array}\right\} \tag{192}$$

Because of eqns. (192), we see that eqns. (171) become

$$\left.\begin{array}{ll} \dfrac{\partial a_1}{\partial\lambda} = u(t)[v'(\lambda)\cos\lambda - v(\lambda)\sin\lambda] & \text{(a)} \\ \dfrac{\partial a_2}{\partial\lambda} = u(t)[v'(\lambda)\sin\lambda + v(\lambda)\cos\lambda], & \text{(b)} \end{array}\right\} \tag{193}$$

so that eqn. (172) becomes

$$a = u[v^2 + v'^2]^{1/2}, \tag{194}$$

and then

$$\frac{\partial}{c\partial t}\ln a = \frac{\dot{u}}{cu}. \tag{195}$$

Also, eqns. (174) and (175) become

$$\left.\begin{array}{ll} \dfrac{\partial}{\partial\lambda}\ln a_1 = \dfrac{v'}{v} - \tan\lambda & \text{(a)} \\ \dfrac{\partial}{\partial\lambda}\ln a_2 = \dfrac{v'}{v} + \cot\lambda & \text{(b)} \end{array}\right\} \tag{196}$$

and

$$\left.\begin{array}{ll} \dfrac{\partial}{c\partial t}\ln a_1 = \dfrac{\dot{u}}{cu} & \text{(a)} \\ \dfrac{\partial}{c\partial \text{t}}\ln a_2 = \dfrac{\dot{u}}{cu}. & \text{(b)} \end{array}\right\} \tag{197}$$

In addition

$$\left.\begin{array}{ll} \dfrac{\partial^2}{\partial\lambda c\partial t}\ln a_1 = 0 & \text{(a)} \\ \dfrac{\partial^2}{\partial\lambda \text{c}\partial \text{t}}\ln a_2 = 0 & \text{(b)} \end{array}\right\} \tag{198}$$

and, also,

$$\frac{\partial^2}{\partial\lambda c\partial t}\ln a = 0. \tag{199}$$

Now, setting

$$\left.\begin{aligned} a_1\frac{\partial\omega_1}{c\partial\lambda} &\equiv \Omega_1 \qquad \text{(a)} \\ a_2\frac{\partial\omega_2}{c\partial\lambda} &\equiv \Omega_2, \qquad \text{(b)} \end{aligned}\right\} \tag{200}$$

the {12} Einstein equation can be written

$$-\Omega_1\frac{\dot u}{cu}+\frac{\partial\Omega_1}{c\partial t}+2\Omega_1\frac{\dot u}{cu}+\Omega_1\frac{\dot u}{cu}=0, \tag{201}$$

which gives, upon integration,

$$\Omega_1(\lambda,t)=\frac{\chi(\lambda)}{u^2(t)}, \tag{202}$$

with $\chi(\lambda)$ an arbitrary function of $\lambda$.

Similarly, the {13} Einstein equation gives

$$\Omega_2(\lambda,t)=\frac{\psi(\lambda)}{u^2(t)}, \tag{203}$$

where $\psi(\lambda)$ is another arbitrary function of $\lambda$.

β′) The {02}, {03}, {23} Einstein equations

From eqn. (200a), namely

$$\Omega_1(\lambda,t)=a_1(\lambda,t)\frac{\partial\omega_1(\lambda,t)}{c\partial\lambda}, \tag{204}$$

after substitution of $\Omega_1$ and $a_1$ from eqns. (202) and (192a) respectively, and integration with respect to $\lambda$, I obtain

$$\int\frac{\chi(\lambda)}{v(\lambda)\cos\lambda}d\lambda = u^3(t)\left[\frac{\omega_1(\lambda,t)}{c}+\frac{f(t)}{c}\right], \tag{205}$$

where f(t) is an arbitrary function. Then differentiating eqn. (205) with respect to t, and since f is arbitrary, I get

$$\omega_1(\lambda,t)=\frac{c\xi(\lambda)}{u^3(t)}, \tag{206}$$

with $\xi(\lambda)$ an arbitrary function. Thus eqn. (205) gives

$$\int\frac{\chi(\lambda)}{v(\lambda)\cos\lambda}d\lambda=\xi(\lambda)+\alpha, \tag{207}$$

where $\alpha$ is a constant. Differentiating (with respect to $\lambda$), I get

$$\frac{\chi(\lambda)}{v(\lambda)\cos\lambda}=\xi'(\lambda), \tag{208}$$

from which I find

$$\chi(\lambda)=v(\lambda)\cos\lambda\xi'(\lambda), \tag{209}$$

and, differentiating the latter,

$$\chi'(\lambda)=v'(\lambda)\cos\lambda\xi'(\lambda)-v(\lambda)\sin\lambda\xi'(\lambda)+v(\lambda)\cos\lambda\xi''(\lambda). \tag{210}$$

Similarly, from eqn. (200b), I obtain

$$\omega_2(\lambda,t)=\frac{c\eta(\lambda)}{u^3(t)}, \tag{211}$$

with $\eta(\lambda)$ an arbitrary function, and then

$$\psi(\lambda)=v(\lambda)\sin\lambda\eta'(\lambda), \tag{212}$$

$$\psi'(\lambda)=v'(\lambda)\sin\lambda\eta'(\lambda)+v(\lambda)\cos\lambda\eta'(\lambda)+v(\lambda)\sin\lambda\eta''(\lambda). \tag{213}$$

The {23} Einstein equation yields

$$\frac{k(p+\varepsilon)}{1-(a_1\omega_1/c)^2-(a_2\omega_2/c)^2}=-\frac{1}{2}u^{-2}\left[v^2+v'^2\right]^{-1}\frac{\xi'}{\xi}\frac{\eta'}{\eta}, \tag{214}$$

where

$$k\equiv 8\pi\,\mathrm{G}/c^4. \tag{215}$$

Now, the {02} Einstein equation can be written as

$$\frac{1}{2}\left[v^2+v'^2\right]^{-1}\left\{-\frac{1}{2}\chi\frac{\left[v^2+v'^2\right]'}{\left[v^2+v'^2\right]}+\chi'+3\chi\frac{v'}{v}-\chi(2\tan\lambda-\cot\lambda)\right\}=$$

$$=u^4\frac{a_1\omega_1/c}{1-(a_1\omega_1/c)^2-(a_2\omega_2/c)^2}k(p+\varepsilon). \tag{216}$$

This gives finally

$$\frac{1}{2}\left[v^2+v'^2\right]^{-1}u^{-2}\left\{-\frac{1}{2}\frac{\xi'}{\xi}\frac{\left[v^2+v'^2\right]'}{\left[v^2+v'^2\right]}+4\frac{v'}{v}\frac{\xi'}{\xi}-3\tan\lambda\frac{\xi'}{\xi}+\frac{\xi''}{\xi}+\frac{\xi'}{\xi}\cot\lambda\right\}=$$

$$=\frac{k(p+\varepsilon)}{1-(a_1\omega_1/c)^2-(a_2\omega_2/c)^2}. \tag{217}$$

Thus, from eqns. (217) and (214), I have

$$\frac{1}{2}\left[v^2+v'^2\right]^{-1}u^{-2}\left\{-\frac{1}{2}\frac{\xi'}{\xi}\frac{\left[v^2+v'^2\right]'}{\left[v^2+v'^2\right]}+4\frac{v'}{v}\frac{\xi'}{\xi}-3\tan\lambda\frac{\xi'}{\xi}+\frac{\xi''}{\xi}+\frac{\xi'}{\xi}\cot\lambda\right\}=$$

$$=-\frac{1}{2}u^{-2}\left[v^2+v'^2\right]^{-1}\frac{\xi'}{\xi}\frac{\eta'}{\eta}, \tag{218}$$

which can be written

$$\frac{d\xi'}{d\lambda}=X(\lambda)\xi', \tag{219}$$

with

$$X(\lambda)\equiv\left(\ln\left[v^2+v'^2\right]^{1/2}\right)'-4(\ln v)'-(\ln\eta)'+3\tan\lambda-\cot\lambda. \tag{220}$$

Integrating eqn. (219), I find

$$\xi'=P\exp\int X(\lambda)d\lambda, \tag{221}$$

where P is a constant. Upon one more integration, the latter gives

$$\xi=P\int e^{\int X(\lambda)d\lambda}d\lambda+const., \tag{222}$$

which yields

$$\xi=P\int\eta^{-1}(\lambda)\frac{\left[v^2+v'^2\right]^{1/2}}{v^4}\left\{\exp\int(3\tan\lambda-\cot\lambda)d\lambda\right\}d\lambda+const.. \tag{223}$$

Performing the inner integration, I am finally left with

$$\xi(\lambda)=P\int\eta^{-1}(\lambda)\frac{\left[v^2+v'^2\right]^{1/2}}{v^4\cos^3\lambda\sin\lambda}d\lambda+const.. \tag{224}$$

Similarly, the {03} Einstein equation, if I take also in mind the {23} Einstein equation, gives finally

$$\eta(\lambda)=Q\int\xi^{-1}(\lambda)\frac{\left[v^2+v'^2\right]^{1/2}}{v^4\sin^3\lambda\cos\lambda}d\lambda+const., \tag{225}$$

where Q is another constant.

If I differentiate (224) and (225) and divide then by parts, I have

$$\frac{\xi'}{\xi} = \frac{P}{Q}\tan^2\lambda\frac{\eta'}{\eta}, \tag{226}$$

while, if I multiply by parts, I have

$$(\xi\xi')(\eta\eta') = PQ\frac{\left[v^2 + v'^2\right]}{v^8\cos^4\lambda\sin^4\lambda}. \tag{227}$$

Thus from the {02}and {03} Einstein equations I get eqns. (224) and (225) respectively, or eqns. (226) and (227), after using the {23} Einstein equation, which has to be also into account. The latter can be written

$$2ku^2\left[v^2 + v'^2\right](p+\varepsilon) + \left\{1 - \left(\frac{v\cos\lambda\xi}{u^2}\right)^2 - \left(\frac{v\sin\lambda\eta}{u^2}\right)^2\right\}\frac{\xi'}{\xi}\frac{\eta'}{\eta} = 0. \tag{228}$$

Now, I proceed to solve the system of simultaneous (ordinary) differential equations (226) & (227) for $\xi$ and $\eta$. Firstly, I find from eqn. (226)

$$\xi = \exp\left\{\frac{P}{Q}\int\tan^2\lambda(\ln\eta)'d\lambda\right\}. \tag{229}$$

If I find $\eta$, then from eqn. (229) I can find $\xi$.

Then, from both eqns. (229) & (227), I obtain for $\eta$ the (ordinary) differential equation

$$\eta\eta' + \frac{P}{Q}\tan^2\lambda(\eta')^2 - \left(\ln\frac{[v^2+v'^2]^{1/2}}{v^4\cos\lambda\sin^3\lambda}\right)'\eta\eta' = 0. \tag{230}$$

Set

$$\eta = \varepsilon\, e^y, \tag{231}$$

where $\varepsilon = \pm 1$. Then eqn. (230) becomes

$$(y')^2 + y'' + \frac{P}{Q}\tan^2\lambda(y')^2 - \left(\ln\frac{[v^2+v'^2]^{1/2}}{v^4\cos\lambda\sin^3\lambda}\right)'y' = 0. \tag{232}$$

Set

$$y' = z. \tag{233}$$

Then eqn. (232) becomes

Eqn. (234) is of the Riccati type (Kappos, 1966), without the “constant” term. Setting

$$z = 1/s, \tag{235}$$

it becomes

$$s' + \left( \ln \frac{[v^2 + v'^2]^{1/2}}{v^4 \cos\lambda \sin^3\lambda} \right)' s = \left( 1 + \frac{P}{Q} \tan^2\lambda \right). \tag{236}$$

This is a linear (differential) equation of the first order. The corresponding homogeneous (differential) equation is

$$s' + \left( \ln \frac{[v^2 + v'^2]^{1/2}}{v^4 \cos\lambda \sin^3\lambda} \right)' s = 0. \tag{237}$$

Solving it, I find

$$s = \Theta \frac{v^4 \cos\lambda \sin^3\lambda}{[v^2 + v'^2]^{1/2}}, \tag{238}$$

where $\Theta$ is a constant. Setting

$$\Theta = \Theta(\lambda), \tag{239}$$

eqn. (236) becomes

$$\Theta'(\lambda) \frac{v^4 \cos\lambda \sin^3\lambda}{[v^2 + v'^2]^{1/2}} = 1 + \frac{P}{Q} \tan^2\lambda. \tag{240}$$

Solving it, and introducing its solution in eqn. (238), I obtain finally

$$s = \frac{\sin^2\lambda}{Q} \frac{v^4 \cos\lambda \sin\lambda}{[v^2 + v'^2]^{1/2}} \int \left( \frac{Q}{\sin^2\lambda} + \frac{P}{\cos^2\lambda} \right) \frac{[v^2 + v'^2]^{1/2}}{v^4 \cos\lambda \sin\lambda} d\lambda. \tag{241}$$

Thus, I find, because of eqn. (235),

$$z = \frac{\dfrac{Q}{\sin^2\lambda} \dfrac{[v^2 + v'^2]^{1/2}}{v^4 \cos\lambda \sin\lambda}}{\displaystyle\int \left( \frac{Q}{\sin^2\lambda} = \frac{P}{\cos^2\lambda} \right) \frac{[v^2 + v'^2]^{1/2}}{v^4 \cos\lambda \sin\lambda} d\lambda}. \tag{242}$$

And, because of eqn. (233),

$$y = \int \frac{\dfrac{Q}{\sin^2\lambda} \dfrac{[v^2 + v'^2]^{1/2}}{v^4 \cos\lambda \sin\lambda}}{\displaystyle\int \left( \frac{Q}{\sin^2\lambda} + \frac{P}{\cos^2\lambda} \right) \frac{[v^2 + v'^2]^{1/2}}{v^4 \cos\lambda \sin\lambda} d\lambda} d\lambda. \tag{243}$$

Finally, because of eqn. (231), I get the solution

$$\eta(\lambda) = \exp\int \frac{\dfrac{Q}{\sin^2\lambda}\,\dfrac{[v^2+v'^2]^{1/2}}{v^4\cos\lambda\sin\lambda}}{\displaystyle\int\left(\frac{Q}{\sin^2\lambda}+\frac{P}{\cos^2\lambda}\right)\frac{[v^2+v'^2]^{1/2}}{v^4\cos\lambda\sin\lambda}\,d\lambda}\,d\lambda. \tag{244}$$

And then eqn. (244) introduced into eqn. (229) gives

$$\xi(\lambda) = \exp\int \frac{\dfrac{P}{\cos^2\lambda}\,\dfrac{[v^2+v'^2]^{1/2}}{v^4\cos\lambda\sin\lambda}}{\displaystyle\int\left(\frac{P}{\cos^2\lambda}+\frac{Q}{\sin^2\lambda}\right)\frac{[v^2+v'^2]^{1/2}}{v^4\cos\lambda\sin\lambda}\,d\lambda}\,d\lambda. \tag{245}$$

γ′) The diagonal Einstein equations

After some manipulations, the {00} equation gives

$$\frac{6}{c^2}u\ddot{u}\left[v^2+v'^2\right]+\frac{v^2}{u^4}\left[\cos^2\lambda\;\xi'^2+\sin^2\lambda\;\eta'^2\right]=\frac{\xi'}{\xi}\frac{\eta'}{\eta}-k(p-\varepsilon)u^2\left[v^2+v'^2\right], \tag{246}$$

the {11} equation gives

$$\frac{2}{c^2}u\ddot{u}\left[v^2+v'^2\right]+\frac{4}{c^2}\dot{u}^2\left[v^2+v'^2\right]+\frac{v^2}{u^4}\left[\cos^2\lambda\,\xi'^2+\sin^2\lambda\,\eta'^2\right]+\frac{\left[v^2+v'^2\right]'}{\left[v^2+v'^2\right]}\left(2\frac{v'}{v}-\tan\lambda+\cot\lambda\right)-$$
$$-4\left(\frac{v''}{v}-1+\left(-\tan\lambda+\cot\lambda\right)\frac{v'}{v}\right)=-k(p-\varepsilon)u^2\left[v^2+v'^2\right], \tag{247}$$

the {22} equation gives

$$\frac{2}{c^2}u\ddot{u}\left[v^2+v'^2\right]+\frac{4}{c^2}\dot{u}^2\left[v^2+v'^2\right]+2\left\{\frac{1}{2}\frac{\left[v^2+v'^2\right]'}{\left[v^2+v'^2\right]}\left(\frac{v'}{v}-\tan\lambda\right)-\frac{v''}{v}-\frac{v'^2}{v^2}+2+\left(3\tan\lambda-\cot\lambda\right)\frac{v'}{v}\right\}-$$
$$-\frac{v^2}{u^4}\cos^2\lambda\,\xi'^2=-\frac{v^2}{u^4}\cos^2\lambda\,\xi^2\frac{\xi'}{\xi}\frac{\eta'}{\eta}-k(p-\varepsilon)u^2\left[v^2+v'^2\right], \tag{248}$$

and the {33} equation gives

$$\frac{2}{c^2}u\ddot{u}\left[v^2+v'^2\right]+\frac{4}{c^2}\dot{u}^2\left[v^2+v'^2\right]+2\left\{\frac{1}{2}\frac{\left[v^2+v'^2\right]'}{\left[v^2+v'^2\right]}\left(\frac{v'}{v}+\cot\lambda\right)-\frac{v''}{v}-\frac{v'^2}{v^2}+2+(\tan\lambda-3\cot\lambda)\frac{v'}{v}\right\}-$$
$$-\frac{v^2}{u^4}\sin^2\lambda\,\eta'^2=-\frac{v^2}{u^4}\sin^2\lambda\,\eta^2\frac{\xi'}{\xi}\frac{\eta'}{\eta}-k(p-\varepsilon)u^2\left[v^2+v'^2\right]. \tag{249}$$

Subtracting eqn. (249) from eqn. (248), we get

$$-\frac{\left[v^2+v'^2\right]'}{\left[v^2+v'^2\right]}(\tan\lambda+\cot\lambda)+4(\tan\lambda+\cot\lambda)\frac{v'}{v}-$$

$$-\frac{v^2}{u^4}\left[\cos^2\lambda\,\xi'^2-\sin^2\lambda\,\eta'^2\right]=-\frac{v^2}{u^4}\left[\cos^2\lambda\,\xi^2-\sin^2\lambda\,\eta^2\right]\frac{\xi'}{\xi}\frac{\eta'}{\eta}. \qquad (250)$$

Differentiating it with respect to time, we obtain

$$\cos^2\lambda\,\xi'^2-\sin^2\lambda\,\eta'^2=\left[\cos^2\lambda\,\xi^2-\sin^2\lambda\,\eta^2\right]\frac{\xi'}{\xi}\frac{\eta'}{\eta}, \qquad (251)$$

from which we find

$$\frac{P^2}{\cos^2\lambda}\xi^2-\frac{Q^2}{\sin^2\lambda}\eta^2=PQ\left[\frac{\xi^2}{\sin^2\lambda}-\frac{\eta^2}{\cos^2\lambda}\right], \qquad (252)$$

or

$$P\left(\frac{P}{\cos^2\lambda}-\frac{Q}{\sin^2\lambda}\right)\xi^2=Q\left(\frac{Q}{\sin^2\lambda}-\frac{P}{\cos^2\lambda}\right)\eta^2, \qquad (253)$$

so that, simplifying, we are left with

$$P\xi^2+Q\eta^2=0. \qquad (254)$$

Then v = v(λ) can be found from eqn. (250) if its terms containing u are neglected. The latter can be done *by virtue* simply of eqn. (254). Thus, we have to solve the equation

$$-\frac{\left[v^2+v'^2\right]'}{\left[v^2+v'^2\right]}(\tan\lambda+\cot\lambda)+4(\tan\lambda+\cot\lambda)\frac{v'}{v}=0, \qquad (255)$$

or

$$\frac{\left[v^2+v'^2\right]'}{\left[v^2+v'^2\right]}=4\frac{v'}{v}, \qquad (256)$$

which reduces to the equation

$$vv'v''-2v'^3-v^2v'=0. \qquad (257)$$

We set

$$v=e^w. \qquad (258)$$

Then eqn. (257) becomes

$$w'w''-w'^3-w'=0. \qquad (259)$$

Set now w′ = x . Then we take from eqn. (259)

$$xx' - x^3 - x = 0. \tag{260}$$

This latter equation gives

$$\int \frac{dx}{x^2 + 1} = \int d\lambda \ , \tag{261}$$

or

$$\lambda + C = \arctan x , \tag{262}$$

where C is a constant of integration, from which we take the solution

$$x = \tan(\lambda + C). \tag{263}$$

From this we find

$$w + G = -\ln\cos(\lambda + C), \tag{264}$$

with G another constant of integration, so that we finally obtain, because of eqn. (258), the solution

$$v = \frac{\Theta}{\cos(\lambda)} , \tag{265}$$

where Θ is a constant ($\Theta = e^{-G}$ ), and we have taken C = 0.

Note that, with C = 0, we first find (cf. eqns. (192))

$$\left.\begin{aligned} & a_1(\lambda,t) = u(t)\Theta, && \text{(a)} \\ & \text{and} \\ & a_2(\lambda,t) = u(t)\Theta\tan\lambda . && \text{(b)} \end{aligned}\right\} \tag{271}$$

Second, the solutions (245) and (244) become

$$\left.\begin{aligned} & \xi(\lambda) = \exp\int \frac{\dfrac{P}{\cos^2\lambda}\cot\lambda}{\displaystyle\int\left(\frac{P}{\cos^2\lambda} + \frac{Q}{\sin^2\lambda}\right)\cot\lambda\, d\lambda} d\lambda , && \text{(a)} \\ & \text{and} \\ & \eta(\lambda) = \exp\int \frac{\dfrac{Q}{\sin^2\lambda}\cot\lambda}{\displaystyle\int\left(\frac{Q}{\sin^2\lambda} + \frac{P}{\cos^2\lambda}\right)\cot\lambda\, d\lambda} d\lambda . && \text{(b)} \end{aligned}\right\} \tag{272}$$

What is left is the determination of u and p-ε. This must be done using the remaining (diagonal) Einstein equations, specifically the {00} and {11} equations, that is eqns. (246) and (247). Eqn. (247) can be written as

$$u^4\left\{\frac{2}{c^2}\left(u\ddot{u}+2\dot{u}^2\right)+k(p-\varepsilon)u^2\right\}=-\cos^2(\lambda)\left[\cos^2\lambda\,\xi'^2+\sin^2\lambda\,\eta'^2\right]. \qquad (273)$$

Eqn. (246) can be written as

$$u^4\left\{\frac{6}{c^2}u\ddot{u}+k(p-\varepsilon)u^2\right\}=\frac{\xi'\eta'}{\xi\eta}\frac{\cos^4(\lambda)}{\Theta^2}u^4-\cos^2(\lambda)\left[\cos^2\lambda\,\xi'^2+\sin^2\lambda\,\eta'^2\right]. \qquad (274)$$

Subtracting eqn. (274) from eqn. (273), we get

$$u^4\left\{-\frac{4}{c^2}u\ddot{u}+\frac{4}{c^2}\dot{u}^2\right\}=-\frac{1}{\Theta^2}\frac{\xi'\eta'}{\xi\eta}\cos^4(\lambda)u^4, \qquad (275)$$

so that we obtain

$$-\frac{4}{c^2}u\ddot{u}+\frac{4}{c^2}\dot{u}^2=-\frac{1}{\Theta^2}\frac{\xi'\eta'}{\xi\eta}\cos^4(\lambda)\,. \qquad (276)$$

We observe that the left hand side of eqn. (276) depends only on t, while the right hand side of the same equation depends only on $\lambda$. Thus both of them must be equal to the same constant, say D. We take therefore the equations

$$-\frac{4}{c^2}u\ddot{u}+\frac{4}{c^2}\dot{u}^2=D \qquad (277)$$

and

$$-\frac{1}{\Theta^2}\frac{\xi'\eta'}{\xi\eta}\cos^4(\lambda)=D. \qquad (278)$$

Note that we can determine D from eqn. (278). Then substituting it in (277) we can solve for u.

We have from eqn. (278), because of eqn. (254),

$$D=-\frac{1}{\Theta^2}\left(\frac{\xi'}{\xi}\right)^2\cos^4(\lambda). \qquad (279)$$

Solving it for $\xi'/\xi$, we find

$$\frac{\xi'}{\xi} = \Theta\sqrt{-D}\,\frac{1}{\cos^2\lambda}. \qquad (280\text{-}1)$$

But we have from eqn. (272a)

$$\frac{\xi'}{\xi} = \frac{\dfrac{P}{\cos^2\lambda}\cot\lambda}{\displaystyle\int\frac{P}{\cos^2\lambda}\cot\lambda\, d\lambda + \int\frac{Q}{\sin^2\lambda}\cot\lambda\, d\lambda}. \qquad (280\text{-}2)$$

Performing the integrations, we find

$$\int\frac{P}{\cos^2\lambda}\cot\lambda\, d\lambda = P\ln\tan\lambda\,, \qquad (280\text{-}3)$$

and

$$\int\frac{Q}{\sin^2\lambda}\cot\lambda\, d\lambda = -\frac{Q}{2}\frac{1}{\sin^2\lambda}\,, \qquad (280\text{-}4)$$

so that eqn. (280-2) becomes

$$\frac{\xi'}{\xi} = \frac{\dfrac{P}{\sin\lambda\cos\lambda}}{P\ln\tan\lambda - \dfrac{Q}{2}\dfrac{1}{\sin^2\lambda}}. \qquad (280\text{-}5)$$

Comparing (280-1) with (280-5) we must get an identity. We thus find the relation

$$\frac{P}{\tan\lambda} = \Theta\sqrt{-D}\left\{P\ln\tan\lambda - \frac{Q}{2}\frac{1}{\sin^2\lambda}\right\}. \qquad (280\text{-}6)$$

The only way for this relation to be an identity, because of the appearance of $\ln\tan\lambda$, is to take P = 0. But then, because of eqn. (254), we also take Q = 0. Thus eqn. (280-2) will give $\xi'/\xi = 0$, which means that $\xi$ = const. Similarly we get $\eta$ = const. (see eqn. (272b)) (and so $\eta'/\eta = 0$ as well). Finally, from eqn. (280-1), we also obtain D = 0, since $\Theta = e^{-G} \neq 0$.

Now, with the value of D found, eqn. (277) becomes very simple, namely

$$\frac{\mathrm{d}}{\mathrm{dt}}\left(\frac{\dot{u}}{u}\right) = 0. \qquad (281)$$

Its solution is trivial, and thus we find

$$u(t) = Ee^{Gt}, \qquad (282)$$

where E and G are two constants of integration.

To find now p-ε, we substitute u by its solution (282) in eqn. (273), and (273) becomes (if we insert u again)

$$u^4\left\{\frac{6}{c^2}G^2u^2+k(p-\varepsilon)u^2\right\}=-\cos^2(\lambda)\left[\cos^2\lambda\,\xi'^2+\sin^2\lambda\,\eta'^2\right]. \tag{283}$$

Differentiating now this equation with respect to t, we obtain the equation

$$36\frac{G^3}{kc^2}+6G(p-\varepsilon)+\frac{d}{dt}(p-\varepsilon)=0. \tag{284}$$

Letting x ≡ p-ε, we have to solve the equation

$$\dot{x}+6Gx=-36G^3/kc^2. \tag{285}$$

This is a linear (ordinary) differential equation of the first order, and its solution is straightforward. We find

$$x(t)=-Ke^{-6Gt}-6G^2/kc^2, \tag{286}$$

with K a constant of integration. Thus

$$p-\varepsilon\ =-K(u/E)^{-6}-6G^2/kc^2. \tag{287}$$

To find now ε and p separately, we have to use an additional equation, the *equation of state.* As it has already been mentioned, in our case it is contained in the field equations. Namely, from the Einstein equations {02}, {03} & {23}, we have used only a combination of {02} & {23}, and a combination of {03} & {23}. Thus one Einstein equation remains to be used, and we take as such the equation {23}. This is eqn. (214), which gives the equation of state

$$p+\varepsilon=0, \tag{288}$$

because as we have found ξ'/ξ = η'/η = 0.

The determination of K remains. We have used all but one of the Einstein equations. Namely, concerning the {22} and {33} equations, that is (248) and (249) respectively, we have used only their combination {22}-{33}. Thus it remains in addition one of them to be examined, or any other combination of them. We take as such the {22} equation, that is (248). It must lead to the determination of K, since we have already found all the other unknowns, if we insert them in it. First, the terms containing $1/u^4$ must give an identity, that is it must

$$\xi'^2 = \xi^2 \frac{\xi'}{\xi}\frac{\eta'}{\eta}, \tag{291}$$

or

$$\frac{\xi'}{\xi} = \frac{\eta'}{\eta}. \tag{292}$$

But this is indeed the case, because of eqn. (254). Second, the terms not containing $[v^2 + v'^2]$ must also give an identity, namely we must have

$$\frac{1}{2}\frac{\left[v^2+v'^2\right]'}{\left[v^2+v'^2\right]}\left(\frac{v'}{v}-\tan\lambda\right)-\frac{v''}{v}-\frac{v'^2}{v^2}+2+(3\tan\lambda-\cot\lambda)\frac{v'}{v}=0 \tag{293}$$

identically. In fact this is the case, as it is easily found if we insert v from eqn. (265). Third, finally, the remaining terms must also give an identity after the determination of K, that is we must have identically

$$\frac{2}{c^2}u\ddot{u}+\frac{4}{c^2}\dot{u}^2=-k(p-\varepsilon)u^2, \tag{294}$$

or, after substitution of u and p-ε from eqns. (282) and (286),

$$\frac{2}{c^2}E^2G^2e^{2Gt}+\frac{4}{c^2}E^2G^2e^{2Gt}=-kE^2e^{2Gt}\left(-Ke^{-6Gt}-6\frac{G^2}{kc^2}\right). \tag{295}$$

But in order for it to be identically satisfied, <u>we must take K = 0.</u> Then eqn. (295) gives

$$\frac{6}{c^2}E^2G^2=\frac{6}{c^2}E^2G^2, \tag{296}$$

which is in fact an identity!

Thus we have finally, instead of eqns. (286) and (287),

$$p-\varepsilon=-6G^2/kc^2. \tag{297}$$

If we take also in mind eqn. (288), we find ε & p separately to be

$$\varepsilon=3G^2/kc^2 \quad \& \quad p=-3G^2/kc^2. \tag{298}$$

## 5. The metric

The form of the metric, as we have seen, is

$$ds^2 = c^2dt^2 - \left[\left(\frac{\partial a_1}{\partial \lambda}\right)^2 + \left(\frac{\partial a_2}{\partial \lambda}\right)^2\right]d\lambda^2 - a_1{}^2(d\varphi_1 - \omega_1 dt)^2 - a_2{}^2(d\varphi_2 - \omega_2 dt)^2. \quad (301)$$

We have found

$$\left.\begin{array}{l} a_1 = u\Theta \\ a_2 = u\Theta \tan\lambda \end{array}\right\} \quad (302)$$

We set

$$u\Theta \equiv U(t), \quad (303)$$

and we perform the transformation

$$\lambda \to z = \tan\lambda. \quad (304)$$

Then eqns. (302) become

$$\left.\begin{array}{l} a_1 = U(t) \\ a_2 = U(t)z \end{array}\right\} \quad (305)$$

From eqns. (305) now we obtain

$$\frac{\partial a_1}{\partial \lambda} = 0, \quad (306)$$

and

$$\frac{\partial a_2}{\partial \lambda}d\lambda = U(t)dz. \quad (307)$$

Thus, the metric (301) becomes

$$ds^2 = c^2dt^2 - U^2(t)\left\{dz^2 + (d\varphi_1 - \omega_1 dt)^2 + z^2(d\varphi_2 - \omega_2 dt)^2\right\}. \quad (308)$$

This metric is comoving, but not "inertial". We will convert it to an "inertial", but <u>not</u> comoving, metric as follows. We have found (cf. eqn. (279)) that

$$\xi'/\xi = 0, \quad (309)$$

so that $\xi' = 0$, and therefore $\xi = \text{const}$. Then eqn. (254) yields also $\eta = \text{const}$. Now, we know that

$$\omega_1 = \frac{c\xi}{u^3(t)} \quad \& \quad \omega_2 = \frac{c\eta}{u^3(t)} \ . \tag{310}$$

Thus, if we use the solution for u (282), we get

$$\omega_1(t) = \frac{c}{E^3}\xi \ e^{-3Gt} \ , \quad \& \tag{311}$$

$$\omega_2(t) = \frac{c}{E^3}\eta \ e^{-3Gt} \ . \tag{312}$$

Now, let us convert the metric (308) to an "inertial" one. Inserting in the binomials $d\varphi_1$-$\omega_1 dt$ and $d\varphi_2$-$\omega_2 dt$ the angular velocities $\omega_1$and $\omega_2$ from eqns. (311) and (312) respectively, we find

$$d\varphi_1 - \omega_1 dt = d\varphi_1 + \frac{c\xi}{3GE^3} d\left(e^{-3Gt}\right) = d\left(\varphi_1 + \frac{c\xi}{3GE^3} e^{-3Gt}\right) \tag{313}$$

and

$$d\varphi_2 - \omega_2 dt = d\varphi_2 + \frac{c\eta}{3GE^3} d\left(e^{-3Gt}\right) = d\left(\varphi_2 + \frac{c\eta}{3GE^3} e^{-3Gt}\right). \tag{314}$$

Performing the coordinate transformation

$$\begin{Bmatrix} t \\ \lambda \\ \varphi_1 \\ \varphi_2 \end{Bmatrix} \rightarrow \begin{Bmatrix} t \\ z = \tan\lambda \\ \phi = \varphi_1 + \dfrac{c\xi}{3GE^3} e^{-3Gt} \\ \Phi = \varphi_2 + \dfrac{c\eta}{3GE^3} e^{-3Gt} \end{Bmatrix} \tag{315}$$

the metric becomes

$$ds^2 = c^2 dt^2 - U^2(t)\left\{dz^2 + d\phi^2 + z^2 d\Phi^2\right\}, \tag{316}$$

with the *scale factor* U(t) given by

$$U(t) = \Theta E e^{Gt}. \tag{317}$$

This is an "inertial" but <u>not comoving</u> metric, which resembles the Robertson-Walker metric with flat space, but it is of course not the flat Friedmann solution since U(t) and $\varepsilon$ are not given by their Friedmann values. If we set

$$d\sigma \equiv dz^2 + d\phi^2 + z^2 d\Phi^2, \tag{318}$$

then the *space* line element is given by

$$dl = U(t)d\sigma \ , \tag{319}$$

so that the *space-time* line element is given by

$$ds^2 = c^2 dt^2 - dl^2. \tag{320}$$

Note that the *space*, because of the expression (318) for dσ, is flat, while the overall *space-time,* despite formula (320), is curved (in the same fashion as it happens in the "flat" Robertson-Walker case) because right of the scale factor U(t) in eqn. (319).

We observe that the "inertial" system of coordinates found has diagonal metric and $g_{00}$ = 1, that is it is a *synchronous* system of coordinates, as we call it. Thus the proper time τ of it is simply the time coordinate t. On the contrary, the proper time of the comoving system of coordinates, taken from the metric (301) if we collect together all the coefficients of $c^2dt^2$, which gives

$$g_{00} = 1 - U^2(t)\frac{\xi^2}{u^6(t)} - U^2(t)z^2\frac{\eta^2}{u^6(t)}, \tag{321}$$

if we use the well known relation

$$d\tau = dt\sqrt{g_{00}} \ , \tag{322}$$

is of course differrent, and its differential is given by

$$d\tau = dt\sqrt{1 - \frac{\Theta^2}{E^4}e^{-4Gt}\left(\xi^2 + \eta^2 z^2\right)} \ . \tag{323}$$

Note that, in order now to take τ, we have to integrate the relation (323).

## 6. Discussion

First of all we have to remark that the age of the Universe described by the metric (308), or (316), is infinite. This is the case because of the exponential form of the scale factor U(t), see (317). But, nevertheless, we can define a new time T which is zero fot t = - infinity. The age To of the Universe is then finite, and very close to the one determined by the observations (Chaliasos, 2006b).

We have to also remark that *there is no "beginning" at the singularity T = 0.* This is the case because *the scale factor U remains having a meaning, and thus the new time T can now be defined, even for T<0* (Chaliasos, 2006b). This is not the case for the FRW Universe, because in that model the scale factor *a* varies as the square root of t(FRW) near t(FRW) = 0, and thus it has no meaning for negative time, which in this way cannot be defined for t(FRW) < 0.

Note also, that, because of the first of the equations (298), a *steady state* of the Universe is obtained. This reminds us the *perfect cosmological principle* underlying the model of *continuous creation* of Bondi-Gold-Hoyle (Contopoulos & Kotsakis, 1987). Also note that this means that our model is *homogeneous,* and thus it does not contradict the observed *homogeneity* of our Universe. It is nevertheless *anisotropic,* as we will see (Chaliasos, 2006b).

It is to be noted that the above steady state does not contradict the *Ryle effect,* eventhough our proposed model of the Universe does not agree with the big-bang model, since it does not predict a superdense state of the Universe at T = 0. On the contrary, in an unexpected way, it right *explains it,* as we will see later on (Chaliasos, 2006b). This is not the case with the NewtonnianUniverse, and even with the FRW Universe.

From the form of the scale factor (317), we also see that our model of the Universe expands, with the Hubble parameter given by the *constant* G, resembling, because of the exponential, the *de Sitter* Universe, eventhough it is *not* empty. But not only this. From the *exponential* form of the scale factor we conclude that the expansion is *accelerating,* a revolutionary fact that was discovered just lately (Glanz, 1998; Perlmutter *et al,* 1998). And this acceleration follows in a completely *physical* way, *not* demanding the introduction of a cosmological constant Λ. This would imply the presence of an ambiguous "dark energy" component in the Universe, which would result in an extraordinary value of Λ of the order of Ω(Λ0) ~ 10^120 (Carrol, Press, & Turner, 1992), which is far from observed, the value of Λ compatible with the observations being very small.

We mentioned above that in the Universe obtained we could not have a superdense state at the"big-bang". Nevertheless all wavelengths are found to be proportional to the scale factor. Thus they should tend to zero at the "big-bang". But then, by Wien' s law, the *temperature* would be *infinite* at the "big-bang".

Concerning the equation of state, we *found* it to be given by p + ε = 0. Nevertheless, we did not use any cosmological constant to find this. Just the rotation resulted *physically* in it. Had we no rotation, the FRW Universe would be obtained and then only a non-vanishing cosmological constant would be necessary to find this equation of state. But we already have stated that this obligation to take Λ into account would contradict the observations.

Finally let us return to our original motivation behind this new cosmological solution of Einstein' s field equations. The motivation consisted of a possible explanation of the structural pattern of spiral galaxies. We will see with satisfaction (Chaliasos, 2006a) that our effort was completely justified.

## 7. Conclusion

Ending this work, we cannot help admiring how trivial the final solution is, despite the over-complexity of the original equations. We have also to note that not only a rotation of the Universe is possible, leading to a natural explanation of the structure of spiral galaxies, and thus justifying our original motivation, but also, in an unexpected way, it leads to a physical explanation of the confusing "acceleration" of the Universe discovered lately (Glanz, 1998; Perlmutter *et al,* 1998).

**Acknowledgements**

I wish here to express my gratitude to my instructor at The University of Chicago Prof. S. Chandrasekhar (deceased) for teaching me the technics of obtaining the components of the Riemann tensor via the Cartan equations of structure in particular, shortening at most the procedure.

I also wish to thank my instructor and supervisor at the University of Athens Prof. G. Contopoulos (Academician) for his valuable comments on this long work.

## Appendix I: The simply rotating universe

I started with the Robertson-Walker metric in the form (Landau & Lifshitz, 1975)

$$ds^2 = c^2dt^2 - a^2(t)\{d\chi^2 + \Sigma^2(\chi)[d\theta^2 + \sin^2\theta d\phi^2]\}, \qquad (1)$$

where χ, θ, φ are hyperspherical coordinates, a(t) is the scale factor and Σ(χ) is given by

$$\Sigma(\chi) = \begin{cases} \sin\chi & \text{for closed universe,} \\ \chi & \text{for flat universe,} \\ \sinh\chi & \text{for open universe.} \end{cases} \qquad (2)$$

I took Σ(χ) = sinχ, which means that I examined the closed model as representative of all three cases included in eqn. (2). I tried to get a rotating solution of Einstein´s equations from the metric (1) by allowing dφ to be substituted by dφ – ωdt,[(*)] in the same manner as Chandrasekhar did in the 2nd Chapter of his book (Chandrasekhar, 1992). I then would obtain a solution of Einstein´s equation which would generalize the Robertson-Walker metric (1) when rotation around the axis of φ was taken into account. The form of the metric would then be

$$ds^2 = c^2dt^2 - a^2(t)\{d\chi^2 + \sin^2\chi[d\theta^2 + \sin^2\theta(d\phi - \omega dt)^2]\}, \tag{3}$$

where the angular velocity of rotation ω would be in general a function of the world time t. The metric (3) then would give the metric (1) for ω = 0, as we wanted. Thus, what we had to do was to calculate the appropriate quantities (that is the quantities entering in Einstein′s equations, i.e. $R_{ik}$, $T_{ik}$, …) from eqn. (3), and then substitute them in Einstein′s equations, demanding them to be satisfied for a(t) and ω(t) to be determined in this way (i.e. in order for the Einstein equations to be satisfied).

---

(*) The fact that the metric is *axisymmetric* is guaranteed by the absence of φ-dependence in the metric coefficients. The fact that the space points φ and φ+dφ (it is supposed that χ=const. & θ=const.), in the coordinate system in which the metric (1) holds, can be assumed to refer to one and the same time (t or t+dt) is guaranteed by the fact that the coordinate frame is *synchronous.* But if the matter (or the comoving frame) *rotates* around the axis of φ with angular velocity ω, we have to replace dφ by dφ´=dφ-ωdt in eqn. (1) (obtaining eqn. (3) in this way), in order for φ´ and φ´+dφ´ to refer again to one and the same time, namely t+dt. This is the case, because the coordinate frame is no longer synchronous. In other words, we must *subtract* from dφ the angle ωdt through which the matter (which means the comoving frame) has been rotated in the meanwhile.

I first found the metric tensor appropriate for the metric (3), and, via this tensor, I found the necessary Christoffel symbols. From these symbols, I found the components of the Ricci tensor. Then I took the energy-momentum tensor appropriate to a comoving perfect fluid, and thus I was ready to form the Einstein equations to be solved.

The Einstein equations in mixed form read

$$R_i^{\ k} - (1/2)\delta_i^{\ k}R = \left(8\pi G/c^4\right)T_i^{\ k}. \tag{4}$$

Those of them which were identically satisfied were the $({}_0{}^0)$, $({}_1{}^1)$, $({}_2{}^2)$, $({}_3{}^3)$ and $({}_0{}^3)$ & $({}_3{}^0)$ equations. For the $({}_0{}^0)$ equation we found

$$\frac{3}{a^2}\left(\dot{a}^2 + c^2\right) = \frac{8\pi G}{c^2}\varepsilon. \tag{5}$$

But this is merely one of the two (independent) Einstein equations for the *Robertson-Walker* case. For our $({}_1{}^1)$, $({}_2{}^2)$, $({}_3{}^3)$ Einstein equations we found the same equation, namely

$$-\frac{2\ddot{a}}{c^2a} - \frac{\dot{a}^2}{c^2a^2} - \frac{1}{a^2} = \frac{8\pi G}{c^4}p. \tag{6}$$

But this is now merely the other Einstein equation for the *Robertson-Walker* case again.

Thus we expected that ω = 0. In fact, the $({}_0{}^3)$ Einstein equation reduced to

$$\left(-\frac{2\ddot{a}}{c^2 a}+\frac{2\dot{a}^2}{c^2 a^2}+\frac{2}{a^2}\right)\frac{\omega}{c}=0, \quad \text{or} \tag{7}$$

$$\frac{8\pi G}{c^4}(p+\varepsilon)\frac{\omega}{c}=0, \tag{7'}$$

by virtue of (5) & (6), which gives just ω = 0! The same happened if we considered the ($_3^0$) Einstein equation, which reduced to

$$0=\frac{8\pi G}{c^4}(p+\varepsilon)a^2\frac{\omega}{c}\sin^2\chi\sin^2\theta\left(1-\frac{a^2\omega^2}{c^2}\sin^2\chi\sin^2\theta\right)^{-1}, \tag{8}$$

which gives again ω = 0!

It seemed that the Robertson-Walker line element does not admit of a generalization comprising rotation, unless we postulate the equation of state p + ε = 0.

## Appendix II: The doubly rotating universe

I started again with the Robertson-Walker line element, especially the spherical type, which I considered as representative of all three possible types: spherical, flat, and hyperbolic. As we know, our spherical space can be imagined as the three-dimensional "surface" of a four-dimensional "sphere". In other words I considered our three-dimensional space as embedded in a fictitious Euclidean four-dimensional space in such a way that the three dimensional space was a three-dimensional spherical hypersurface centered at the origin.

Let $x_1$, $x_2$, $x_3$, $x_4$ be the coordinates in the four-dimensional space. Obviously, we can leave the above three-dimensional spherical hypersurface, of radius say a, *invariant* by performing a rotation on the ($x_1$, $x_2$) plane, which does not affect the coordinates $x_3$ & $x_4$, and at the same time performing a rotation on the ($x_3$, $x_4$) plane, which does not affect the coordinates $x_1$ & $x_2$. Suppose that the first rotation is through an angle $\varphi_1$, and the second one through an angle $\varphi_2$. Let the radius vector on the ($x_1$, $x_2$) plane have a length $a_1$, and the radius vector on the ($x_3$, $x_4$) plane have a length $a_2$. Then the above *two* rotations will be given by

$$\left.\begin{aligned} x_1 &= a_1 \cos\phi_1 \\ x_2 &= a_1 \sin\phi_1 \end{aligned}\right\} \text{(a)} \quad \& \quad \left.\begin{aligned} x_3 &= a_2 \cos\phi_2 \\ x_4 &= a_2 \sin\phi_2 \end{aligned}\right\} \text{(b)} \qquad (9)$$

We will then have evidently

$$a^2 = x_1^{\ 2} + x_2^{\ 2} + x_3^{\ 2} + x_4^{\ 2} = a_1^{\ 2} + a_2^{\ 2}. \qquad (10)$$

Thus, we can also set

$$\left.\begin{aligned} a_1 &= a\cos\theta \\ a_2 &= a\sin\theta \end{aligned}\right\} \qquad (11)$$

In this way, the overall transformation will be

$$\left.\begin{aligned} x_1 &= a\cos\theta\,\cos\phi_1 \\ x_2 &= a\cos\theta\,\sin\phi_1 \end{aligned}\right\} \text{(a)} \quad \& \quad \left.\begin{aligned} x_3 &= a\sin\theta\,\cos\phi_2 \\ x_4 &= a\sin\theta\,\sin\phi_2 \end{aligned}\right\} \text{(b)} \qquad (12)$$

We can now consider the transformation (12) as giving the transformation of *coordinates*

$$(x_1, x_2, x_3, x_4) \to (a, \theta, \phi_1, \phi_2). \qquad (13)$$

Our hypersurface will be given by a=const., so that its (three) coordinates will be $\theta$, $\varphi_1$, $\varphi_2$.

Let dL mean the line element of the fictitious Euclidean four-dimensional space. It will obviously be given by

$$dL^2 = dx_1^{\ 2} + dx_2^{\ 2} + dx_3^{\ 2} + dx_4^{\ 2}. \tag{14}$$

Taking into account the relations (12), we can express the differentials on the right-hand side of formula (14) through the new coordinates. Adding then up their squares, we are left with

$$dL^2 = da^2 + a^2 \cos^2\theta d\phi_1^{\ 2} + a^2 \sin^2\theta d\phi_2^{\ 2} + a^2 d\theta^2. \tag{15}$$

Thus the line element on our hypersurface (a = const.) will be

$$dl^2 = a^2(d\theta^2 + \cos^2\theta d\phi_1^{\ 2} + \sin^2\theta d\phi_2^{\ 2}). \tag{16}$$

Then, if we use the world time t, we will have for the line element of the four-dimensional *space-time,* that is the Robertson-Walker line element,

$$ds^2 = c^2 dt^2 - dl^2, \tag{17}$$

with $dl^2$ given by eqn. (16)

The important point on the Robertson-Walker metric in the form (17) is that the new coordinates admit of *two* rotations: one through the angle $\varphi_1$ and the other through the angle $\varphi_2$. Thus, in order to find the metric of this *doubly* rotating Universe, we have to do what we did in the previous section, after making the *two* substitutions

$$\left.\begin{aligned} d\phi_1 &\rightarrow d\phi_1 - \omega_1 dt \\ d\phi_2 &\rightarrow d\phi_2 - \omega_2 dt \end{aligned}\right\} \tag{18}$$

leading finally to the form of the metric

$$ds^2 = c^2 dt^2 - a^2\{d\theta^2 + \cos^2\theta(d\phi_1 - \omega_1 dt)^2 + \sin^2\theta(d\phi_2 - \omega_2 dt)^2\}. \tag{19}$$

I first found the metric tensor appropriate to the metric (19).

Then I wrote the line element (19) as

$$ds^2 = \eta_{ab}\left(e^{(a)}{}_i dx^i\right)\left(e^{(b)}{}_k dx^k\right), \tag{20}$$

that is in tetrad form (Landau & Lifshitz, 1975; Chandrasekhar, 1992). I determined the appropriate tetrad by finding its′ basis vectors′ (tensor) components, corresponding to the constant matrix

Alternatively, we can define the quantities inside the parentheses in eqn. (20) as one-forms (Chandrasekhar, 1992; Lovelock & Rund, 1975) appropriate to the chosen tetrad

$$\omega^a = e^{(a)}{}_i dx^i . \tag{22}$$

That is, analytically, we define the one-forms

$$\left.\begin{aligned} \omega^0 &= cdt \\ \omega^1 &= ad\theta \\ \omega^2 &= a\cos\theta(d\phi_1 - \omega_1 dt) \\ \omega^3 &= a\sin\theta(d\phi_2 - \omega_2 dt) \end{aligned}\right\} \tag{23}$$

I first determined the tetrad components of the Ricci tensor via the Cartan equations of structure (Chandrasekhar, 1992; Lovelock & Rund, 1975). I confirmed then the correctness of my results by computing them via the λ-symbols of tetrad formalism (Landau & Lifshitz, 1975; Chandrasekhar, 1992). Next, I expressed the energy-momentum tensor in tetrad form, appropriate to a comoving perfect fluid. Thus I could form the Einstein equations in tetrad form, as

$$R^{(a)(b)} - \frac{1}{2}\eta^{ab}R = \frac{8\pi G}{c^4}T^{(a)(b)} \tag{24}$$

I found that the (01), (12),(13) equations, as well as the (23) one, are merely identities. The (02) & (03) equations give, respectively,

$$0 = -\frac{a(\omega_1/c)\cos\theta}{g_{00}}(p+\varepsilon), \text{ and} \tag{25}$$

$$0 = -\frac{a(\omega_2/c)\sin\theta}{g_{00}}(p+\varepsilon), \tag{26}$$

where p is the pressure and ε the energy density of the cosmic fluid. We thus obtain from these equations, respectively,

Thus, there is again no rotation at all, if we do not want to postulate the equation of state p + ε = 0.

The (00) equation on the one hand, and the coinciding (11), (22), (33) equations on the other hand, give then the two independent Robertson-Walker equations, as expected.

**Appendix III:** *The curvature tensor via Cartan´s equations of structure*

My purpose is now to first find the tetrad components of the curvature tensor via the Cartan equations of structure (Chandrasekhar,1992; Lovelock & Rund, 1975), in order to introduce them into the Einstein equations.

I write again the equations (48) defining the one-forms $\omega^a$. These are

$$\omega^0 = cdt$$

$$\omega^1 = \left[\left(\frac{\partial a_1}{\partial\lambda}\right)^2 + \left(\frac{\partial a_2}{\partial\lambda}\right)^2\right]^{1/2} d\lambda$$

$$\omega^2 = a_1(d\phi_1 - \omega_1 dt)$$

$$\omega^3 = a_2(d\phi_2 - \omega_2 dt) \qquad (68)$$

From these equations I can find the differentials of the coordinates cdt, dλ, $d\varphi_1$, $d\varphi_2$, needed in the sequel, as functions of the one-forms $\omega^a$. I find

$$cdt = \omega^0$$

$$d\lambda = \left[\left(\frac{\partial a_1}{\partial\lambda}\right)^2 + \left(\frac{\partial a_2}{\partial\lambda}\right)^2\right]^{-1/2} \omega^1$$

$$d\phi_1 = \frac{\omega_1}{c}\omega^0 + \frac{1}{a_1}\omega^2$$

$$d\phi_2 = \frac{\omega_2}{c}\omega^0 + \frac{1}{a_2}\omega^3. \qquad (69)$$

Taking the exterior derivatives of the one-forms $\omega^a$ given by eqns.(68), and using eqns. (69), I obtain

$$d\omega^0 = 0$$

$$d\omega^1 = \frac{\partial}{c\partial t}\ln\left[\left(\frac{\partial a_1}{\partial\lambda}\right)^2 + \left(\frac{\partial a_2}{\partial\lambda}\right)^2\right]^{1/2}\omega^0\wedge\omega^1$$

$$d\omega^2 = a_1\frac{\partial\omega_1}{c\partial\lambda}\left[\left(\frac{\partial a_1}{\partial\lambda}\right)^2 + \left(\frac{\partial a_2}{\partial\lambda}\right)^2\right]^{-1/2}\omega^0\wedge\omega^1 +$$

$$+\frac{\partial}{c\partial t}\ln a_1\omega^0\wedge\omega^2 +$$

$$+\left(\frac{\partial}{\partial\lambda}\ln a_1\right)\left[\left(\frac{\partial a_1}{\partial\lambda}\right)^2 + \left(\frac{\partial a_2}{\partial\lambda}\right)^2\right]^{-1/2}\omega^1\wedge\omega^2$$

$$d\omega^3 = a_2\frac{\partial\omega_2}{c\partial\lambda}\left[\left(\frac{\partial a_1}{\partial\lambda}\right)^2 + \left(\frac{\partial a_2}{\partial\lambda}\right)^2\right]^{-1/2}\omega^0\wedge\omega^1 +$$

$$+\frac{\partial}{c\partial t}\ln a_2\omega^0\wedge\omega^3 +$$

$$+\left(\frac{\partial}{\partial\lambda}\ln a_2\right)\left[\left(\frac{\partial a_1}{\partial\lambda}\right)^2 + \left(\frac{\partial a_2}{\partial\lambda}\right)^2\right]^{-1/2}\omega^1\wedge\omega^3. \qquad (70)$$

Now knowing $\omega^b$ (from (68)) and $d\omega^a$ (from (70)), I insert them into Cartan´s 1st equation of structure (Chandrasekhar, 1992; Lovelock & Rund; 1975) in the case that torsion is zero, reading

$$d\omega^a + \omega^a{}_b\wedge\omega^b = 0 \quad \text{(Cartan I)}, \qquad (71)$$

in order to find the one-forms $\omega^a{}_b$ , needed in the sequel. I get for the non-vanishing of them

$$\omega^0{}_1 = \frac{\partial}{c\partial t}\ln[\ ]^{1/2}\omega^1 + \frac{1}{2}[\ ]^{-1/2}\left(a_1\frac{\partial\omega_1}{c\partial\lambda}\omega^2 + a_2\frac{\partial\omega_2}{c\partial\lambda}\omega^3\right)$$

$$\omega^0{}_2 = \frac{1}{2}[\ ]^{-1/2}a_1\frac{\partial\omega_1}{c\partial\lambda}\omega^1 + \frac{\partial}{c\partial t}\ln a_1\omega^2$$

$$\omega^0{}_3 = \frac{1}{2}[\ ]^{-1/2}a_2\frac{\partial\omega_2}{c\partial\lambda}\omega^1 + \frac{\partial}{c\partial t}\ln a_2\omega^3$$

$$\omega^1{}_2 = \frac{1}{2}[\ ]^{-1/2}a_1\frac{\partial\omega_1}{c\partial\lambda}\omega^0 - \left(\frac{\partial}{\partial\lambda}\ln a_1\right)[\ ]^{-1/2}\omega^2$$

$$\omega^1{}_3 = \frac{1}{2}[\ ]^{-1/2}a_2\frac{\partial\omega_2}{c\partial\lambda}\omega^0 - \left(\frac{\partial}{\partial\lambda}\ln a_2\right)[\ ]^{-1/2}\omega^3. \qquad (72)$$

The Riemann tensor sought though appears into Cartan´s 2nd equation of structure, and it can be extracted from this equation. But the exterior derivatives of the one-forms $\omega^i{}_j$ found above appear also in Cartan´s 2nd equation of structure. Thus we have to compute them from (72) and insert them into Cartan´s 2nd equation of structure in order to find the (tetrad) components of the Riemann tensor. I find then

$$\begin{aligned} d\omega^0{}_1 = &\left\{\frac{\partial^2}{c^2\partial t^2}\ln[\ ]^{1/2} + \left(\frac{\partial}{c\partial t}\ln[\ ]^{1/2}\right)^2 + \frac{1}{2}[\ ]^{-1}\left[\left(a_1\frac{\partial\omega_1}{c\partial\lambda}\right)^2 + \left(a_2\frac{\partial\omega_2}{c\partial\lambda}\right)^2\right]\right\}\omega^0\wedge\omega^1 + \\ &+\frac{1}{2}[\ ]^{-1/2}\left\{-\left(a_1\frac{\partial\omega_1}{c\partial\lambda}\right)\frac{\partial}{c\partial t}\ln[\ ]^{1/2} + \frac{\partial}{c\partial t}\left(a_1\frac{\partial\omega_1}{c\partial\lambda}\right) + \left(a_1\frac{\partial\omega_1}{c\partial\lambda}\right)\frac{\partial}{c\partial t}\ln a_1\right\}\omega^0\wedge\omega^2 + \\ &+\frac{1}{2}[\ ]^{-1/2}\left\{-\left(a_2\frac{\partial\omega_2}{c\partial\lambda}\right)\frac{\partial}{c\partial t}\ln[\ ]^{1/2} + \frac{\partial}{c\partial t}\left(a_2\frac{\partial\omega_2}{c\partial\lambda}\right) + \left(a_2\frac{\partial\omega_2}{c\partial\lambda}\right)\frac{\partial}{c\partial t}\ln a_2\right\}\omega^0\wedge\omega^3 + \\ &+\frac{1}{2}[\ ]^{-1}\left\{-\left(a_1\frac{\partial\omega_1}{c\partial\lambda}\right)\frac{\partial}{\partial\lambda}\ln[\ ]^{1/2} + \frac{\partial}{\partial\lambda}\left(a_1\frac{\partial\omega_1}{c\partial\lambda}\right) + \left(a_1\frac{\partial\omega_1}{c\partial\lambda}\right)\frac{\partial}{\partial\lambda}\ln a_1\right\}\omega^1\wedge\omega^2 + \\ &+\frac{1}{2}[\ ]^{-1}\left\{-\left(a_2\frac{\partial\omega_2}{c\partial\lambda}\right)\frac{\partial}{\partial\lambda}\ln[\ ]^{1/2} + \frac{\partial}{\partial\lambda}\left(a_2\frac{\partial\omega_2}{c\partial\lambda}\right) + \left(a_2\frac{\partial\omega_2}{c\partial\lambda}\right)\frac{\partial}{\partial\lambda}\ln a_2\right\}\omega^1\wedge\omega^3 \end{aligned} \qquad (73)$$

$$d\omega^0{}_2 = \frac{1}{2}[\quad]^{-1/2}\left\{\frac{\partial}{c\partial t}\left(a_1\frac{\partial\omega_1}{c\partial\lambda}\right)+2\left(a_1\frac{\partial\omega_1}{c\partial\lambda}\right)\frac{\partial}{c\partial t}\ln a_1\right\}\omega^0\wedge\omega^1 +$$
$$+\left\{\frac{\partial^2}{c^2\partial t^2}\ln a_1+\left(\frac{\partial}{c\partial t}\ln a_1\right)^2\right\}\omega^0\wedge\omega^2 +$$
$$+[\quad]^{-1/2}\left\{\frac{\partial^2}{\partial\lambda c\partial t}\ln a_1+\left(\frac{\partial}{c\partial t}\ln a_1\right)\left(\frac{\partial}{\partial\lambda}\ln a_1\right)\right\}\omega^1\wedge\omega^2 \tag{74}$$

$$d\omega^0{}_3 = \frac{1}{2}[\quad]^{-1/2}\left\{\frac{\partial}{c\partial t}\left(a_2\frac{\partial\omega_2}{c\partial\lambda}\right)+2\left(a_2\frac{\partial\omega_2}{c\partial\lambda}\right)\frac{\partial}{c\partial t}\ln a_2\right\}\omega^0\wedge\omega^1 +$$
$$+\left\{\frac{\partial^2}{\mathrm{c}^2\partial t^2}\ln a_2+\left(\frac{\partial}{c\partial t}\ln a_2\right)^2\right\}\omega^0\wedge\omega^3 +$$
$$+[\quad]^{-1/2}\left\{\frac{\partial^2}{\partial\lambda c\partial t}\ln a_2+\left(\frac{\partial}{c\partial t}\ln a_2\right)\left(\frac{\partial}{\partial\lambda}\ln a_2\right)\right\}\omega^1\wedge\omega^3 \tag{75}$$

$$d\omega^1{}_2 = \frac{1}{2}[\quad]^{-1}\left\{\left(a_1\frac{\partial\omega_1}{c\partial\lambda}\right)\frac{\partial}{\partial\lambda}\ln[\quad]^{1/2}-\frac{\partial}{\partial\lambda}\left(a_1\frac{\partial\omega_1}{c\partial\lambda}\right)-2\left(a_1\frac{\partial\omega_1}{c\partial\lambda}\right)\frac{\partial}{\partial\lambda}\ln a_1\right\}\omega^0\wedge\omega^1 +$$
$$+[\quad]^{-1/2}\left\{\left(\frac{\partial}{c\partial t}\ln[\quad]^{1/2}\right)\left(\frac{\partial}{\partial\lambda}\ln a_1\right)-\frac{\partial^2}{c\partial t\partial\lambda}\ln a_1-\left(\frac{\partial}{c\partial t}\ln a_1\right)\left(\frac{\partial}{\partial\lambda}\ln a_1\right)\right\}\omega^0\wedge\omega^2 +$$
$$+[\quad]^{-1}\left\{\left(\frac{\partial}{\partial\lambda}\ln[\quad]^{1/2}\right)\left(\frac{\partial}{\partial\lambda}\ln a_1\right)-\frac{\partial^2}{\partial\lambda^2}\ln a_1-\left(\frac{\partial}{\partial\lambda}\ln a_1\right)^2\right\}\omega^1\wedge\omega^2 \tag{76}$$

$$d\omega^1{}_3 = \frac{1}{2}[\;]^{-1}\left\{\left(a_2\frac{\partial\omega_2}{c\partial\lambda}\right)\frac{\partial}{\partial\lambda}\ln[\;]^{1/2} - \frac{\partial}{\partial\lambda}\left(a_2\frac{\partial\omega_2}{c\partial\lambda}\right) - 2\left(a_2\frac{\partial\omega_2}{c\partial\lambda}\right)\frac{\partial}{\partial\lambda}\ln a_2\right\}\omega^0\wedge\omega^1 +$$
$$+[\;]^{-1/2}\left\{\left(\frac{\partial}{c\partial t}\ln[\;]^{1/2}\right)\left(\frac{\partial}{\partial\lambda}\ln a_2\right) - \frac{\partial^2}{c\partial t\partial\lambda}\ln a_2 - \left(\frac{\partial}{c\partial t}\ln a_2\right)\left(\frac{\partial}{\partial\lambda}\ln a_2\right)\right\}\omega^0\wedge\omega^3 +$$
$$+[\;]^{-1}\left\{\left(\frac{\partial}{\partial\lambda}\ln[\;]^{1/2}\right)\left(\frac{\partial}{\partial\lambda}\ln a_2\right) - \frac{\partial^2}{\partial\lambda^2}\ln a_2 - \left(\frac{\partial}{\partial\lambda}\ln a_2\right)^2\right\}\omega^1\wedge\omega^3 \qquad (77)$$

Now, the 2nd Cartan´s equation of structure (Chandrasekhar, 1992; Lovelock & Rund, 1975) reads

$$(1/2)R^i{}_{jkl}\omega^k\wedge\omega^l = d\omega^i{}_j + \omega^i{}_m\wedge\omega^m{}_j \qquad \text{(Cartan II)} \qquad (78)$$

or

$$R^i{}_{jkl}\omega^k\wedge\omega^l = d\omega^i{}_j + \omega^i{}_m\wedge\omega^m{}_j \quad (k<l) \quad \text{(Cartan II)} \qquad (79)$$

where $R^i{}_{jkl}$ are the tetrad components of the Riemann tensor. Since the quantities on the right-hand side have been found, we can substitute them into this equation and then find $R^i{}_{jkl}$ from the left-hand side.

From

$$R^0{}_{1kl}\omega^k\wedge\omega^l = d\omega^0{}_1 + \omega^0{}_m\wedge\omega^m{}_1 \qquad (k<l) \qquad (80)$$

I find

$$R^0{}_{101} = \frac{\partial^2}{c^2\partial t^2}\ln[\;]^{1/2} + \left(\frac{\partial}{c\partial t}\ln[\;]^{1/2}\right)^2 + \frac{3}{4}[\;]^{-1}\left[\left(a_1\frac{\partial\omega_1}{c\partial\lambda}\right)^2 + \left(a_2\frac{\partial\omega_2}{c\partial\lambda}\right)^2\right] \qquad (81)$$

$$R^0{}_{102} = \frac{1}{2}[\;]^{-1/2}\left\{-\left(a_1\frac{\partial\omega_1}{c\partial\lambda}\right)\frac{\partial}{c\partial t}\ln[\;]^{1/2} + \frac{\partial}{c\partial t}\left(a_1\frac{\partial\omega_1}{c\partial\lambda}\right) + 2\left(a_1\frac{\partial\omega_1}{c\partial\lambda}\right)\frac{\partial}{c\partial t}\ln a_1\right\} \qquad (82)$$

$$R^0{}_{103} = \frac{1}{2}[\;]^{-1/2}\left\{-\left(a_2\frac{\partial\omega_2}{c\partial\lambda}\right)\frac{\partial}{c\partial t}\ln[\;]^{1/2} + \frac{\partial}{c\partial t}\left(a_2\frac{\partial\omega_2}{c\partial\lambda}\right) + 2\left(a_2\frac{\partial\omega_2}{c\partial\lambda}\right)\frac{\partial}{c\partial t}\ln a_2\right\} \qquad (83)$$

$$R^0{}_{112} = \frac{1}{2}[\ ]^{-1}\left\{-\left(a_1\frac{\partial\omega_1}{c\partial\lambda}\right)\frac{\partial}{\partial\lambda}\ln[\ ]^{1/2} + \frac{\partial}{\partial\lambda}\left(a_1\frac{\partial\omega_1}{c\partial\lambda}\right) + 2\left(a_1\frac{\partial\omega_1}{c\partial\lambda}\right)\frac{\partial}{\partial\lambda}\ln a_1\right\} \tag{84}$$

$$R^0{}_{113} = \frac{1}{2}[\ ]^{-1}\left\{-\left(a_2\frac{\partial\omega_2}{c\partial\lambda}\right)\frac{\partial}{\partial\lambda}\ln[\ ]^{1/2} + \frac{\partial}{\partial\lambda}\left(a_2\frac{\partial\omega_2}{c\partial\lambda}\right) + 2\left(a_2\frac{\partial\omega_2}{c\partial\lambda}\right)\frac{\partial}{\partial\lambda}\ln a_2\right\}. \tag{85}$$

From

$$R^0{}_{2kl}\omega^k \wedge \omega^l = d\omega^0{}_2 + \omega^0{}_m \wedge \omega^m{}_2 \quad (k < l) \tag{86}$$

I find

$$R^0{}_{201} = \frac{1}{2}[\ ]^{-1/2}\left\{-\left(a_1\frac{\partial\omega_1}{c\partial\lambda}\right)\frac{\partial}{c\partial t}\ln[\ ]^{1/2} + \frac{\partial}{c\partial t}\left(a_1\frac{\partial\omega_1}{c\partial\lambda}\right) + 2\left(a_1\frac{\partial\omega_1}{c\partial\lambda}\right)\frac{\partial}{c\partial t}\ln a_1\right\} \tag{87}$$

$$R^0{}_{202} = \frac{\partial^2}{c^2\partial t^2}\ln a_1 + \left(\frac{\partial}{c\partial t}\ln a_1\right)^2 - \frac{1}{4}[\ ]^{-1}\left(a_1\frac{\partial\omega_1}{c\partial\lambda}\right)^2 \tag{88}$$

$$R^0{}_{203} = -\frac{1}{4}[\ ]^{-1}\left(a_1\frac{\partial\omega_1}{c\partial\lambda}\right)\left(a_2\frac{\partial\omega_2}{c\partial\lambda}\right) \tag{89}$$

$$R^0{}_{212} = [\ ]^{-1/2}\left\{-\left(\frac{\partial}{\partial\lambda}\ln a_1\right)\frac{\partial}{c\partial t}\ln[\ ]^{1/2} + \frac{\partial^2}{\partial\lambda c\partial t}\ln a_1 + \left(\frac{\partial}{c\partial t}\ln a_1\right)\left(\frac{\partial}{\partial\lambda}\ln a_1\right)\right\} \tag{90}$$

$$R^0{}_{223} = \frac{1}{2}[\ ]^{-1}\left(a_2\frac{\partial\omega_2}{c\partial\lambda}\right)\frac{\partial}{\partial\lambda}\ln a_1 \tag{91}$$

From

$$R^0{}_{3kl}\omega^k \wedge \omega^l = d\omega^0{}_3 + \omega^0{}_m \wedge \omega^m{}_3 \quad (k < l) \tag{92}$$

I find

$$R^0{}_{301} = \frac{1}{2}[\ ]^{-1/2}\left\{-\left(a_2\frac{\partial\omega_2}{c\partial\lambda}\right)\frac{\partial}{c\partial t}\ln[\ ]^{1/2} + \frac{\partial}{c\partial t}\left(a_2\frac{\partial\omega_2}{c\partial\lambda}\right) + 2\left(a_2\frac{\partial\omega_2}{c\partial\lambda}\right)\frac{\partial}{c\partial t}\ln a_2\right\} \tag{93}$$

$$R^0{}_{302} = -\frac{1}{4}[\ ]^{-1}\left(a_2\frac{\partial\omega_2}{c\partial\lambda}\right)\left(a_1\frac{\partial\omega_1}{c\partial\lambda}\right) \tag{94}$$

$$R^0{}_{303} = \frac{\partial^2}{c^2\partial t^2}\ln a_2 + \left(\frac{\partial}{c\partial t}\ln a_2\right)^2 - \frac{1}{4}[\ ]^{-1}\left(a_2\frac{\partial\omega_2}{c\partial\lambda}\right)^2 \tag{95}$$

$$R^0{}_{313} = [\ ]^{-1/2}\left\{-\left(\frac{\partial}{\partial\lambda}\ln a_2\right)\frac{\partial}{c\partial t}\ln[\ ]^{1/2} + \frac{\partial^2}{\partial\lambda c\partial t}\ln a_2 + \left(\frac{\partial}{c\partial t}\ln a_2\right)\left(\frac{\partial}{\partial\lambda}\ln a_2\right)\right\} \tag{96}$$

$$R^0{}_{323} = -\frac{1}{2}[\ ]^{-1}\left(a_1\frac{\partial\omega_1}{c\partial\lambda}\right)\frac{\partial}{\partial\lambda}\ln a_2. \tag{97}$$

From

$$R^1{}_{2kl}\omega^k \wedge \omega^l = d\omega^1{}_2 + \omega^1{}_0 \wedge \omega^0{}_2 + \omega^1{}_3 \wedge \omega^3{}_2 \qquad (k < l) \tag{98}$$

I find

$$R^1{}_{201} = \frac{1}{2}[\ ]^{-1}\left\{\left(a_1\frac{\partial\omega_1}{c\partial\lambda}\right)\frac{\partial}{\partial\lambda}\ln[\ ]^{1/2} - \frac{\partial}{\partial\lambda}\left(a_1\frac{\partial\omega_1}{c\partial\lambda}\right) - 2\left(a_1\frac{\partial\omega_1}{c\partial\lambda}\right)\frac{\partial}{\partial\lambda}\ln a_1\right\} \tag{99}$$

$$R^1{}_{202} = [\ ]^{-1/2}\left\{\left(\frac{\partial}{c\partial t}\ln[\ ]^{1/2}\right)\left(\frac{\partial}{\partial\lambda}\ln a_1\right) - \frac{\partial^2}{c\partial t\partial\lambda}\ln a_1 - \left(\frac{\partial}{c\partial t}\ln a_1\right)\left(\frac{\partial}{\partial\lambda}\ln a_1\right)\right\} \tag{100}$$

$$R^1{}_{212} = [\ ]^{-1}\left\{\begin{array}{l}\left(\frac{\partial}{\partial\lambda}\ln[\ ]^{1/2}\right)\left(\frac{\partial}{\partial\lambda}\ln a_1\right) - \frac{\partial^2}{\partial\lambda^2}\ln a_1 - \left(\frac{\partial}{\partial\lambda}\ln a_1\right)^2 + \\ +[\ ]\left(\frac{\partial}{c\partial t}\ln[\ ]^{1/2}\right)\left(\frac{\partial}{c\partial t}\ln a_1\right) - \frac{1}{4}\left(a_1\frac{\partial\omega_1}{c\partial\lambda}\right)^2\end{array}\right\} \tag{101}$$

$$R^1{}_{213} = -\frac{1}{4}[\ ]^{-1}\left(a_1\frac{\partial\omega_1}{c\partial\lambda}\right)\left(a_2\frac{\partial\omega_2}{c\partial\lambda}\right) \tag{102}$$

$$R^1{}_{223} = -\frac{1}{2}[\ ]^{-1/2}\left(a_2\frac{\partial\omega_2}{c\partial\lambda}\right)\frac{\partial}{c\partial t}\ln a_1. \tag{103}$$

Also

I find

$$R^1{}_{301} = \frac{1}{2}[\ ]^{-1}\left\{\left(a_2\frac{\partial\omega_2}{c\partial\lambda}\right)\frac{\partial}{\partial\lambda}\ln[\ ]^{1/2} - \frac{\partial}{\partial\lambda}\left(a_2\frac{\partial\omega_2}{c\partial\lambda}\right) - 2\left(a_2\frac{\partial\omega_2}{c\partial\lambda}\right)\frac{\partial}{\partial\lambda}\ln a_2\right\} \quad (105)$$

$$R^1{}_{303} = [\ ]^{-1/2}\left\{\left(\frac{\partial}{c\partial t}\ln[\ ]^{1/2}\right)\left(\frac{\partial}{\partial\lambda}\ln a_2\right) - \frac{\partial^2}{c\partial t\partial\lambda}\ln a_2 - \left(\frac{\partial}{c\partial t}\ln a_2\right)\left(\frac{\partial}{\partial\lambda}\ln a_2\right)\right\} \quad (106)$$

$$R^1{}_{312} = -\frac{1}{4}[\ ]^{-1}\left(a_2\frac{\partial\omega_2}{c\partial\lambda}\right)\left(a_1\frac{\partial\omega_1}{c\partial\lambda}\right) \quad (107)$$

$$R^1{}_{313} = [\ ]^{-1}\left\{\begin{matrix}\left(\frac{\partial}{\partial\lambda}\ln[\ ]^{1/2}\right)\left(\frac{\partial}{\partial\lambda}\ln a_2\right) - \frac{\partial^2}{\partial\lambda^2}\ln a_2 - \left(\frac{\partial}{\partial\lambda}\ln a_2\right)^2 + \\ +[\ ]\left(\frac{\partial}{c\partial t}\ln[\ ]^{1/2}\right)\left(\frac{\partial}{c\partial t}\ln a_2\right) - \frac{1}{4}\left(a_2\frac{\partial\omega_2}{c\partial\lambda}\right)^2\end{matrix}\right\} \quad (109)$$

$$R^1{}_{323} = \frac{1}{2}[\ ]^{-1/2}\left(a_1\frac{\partial\omega_1}{c\partial\lambda}\right)\frac{\partial}{c\partial t}\ln a_2. \quad (110)$$

And from

$$R^2{}_{3kl}\omega^k \wedge \omega^l = d\omega^2{}_3 + \omega^2{}_0 \wedge \omega^0{}_3 + \omega^2{}_1 \wedge \omega^1{}_3 \quad (k < l) \quad (111)$$

I find

$$R^2{}_{302} = -\frac{1}{2}[\ ]^{-1}\left(a_2\frac{\partial\omega_2}{c\partial\lambda}\right)\left(\frac{\partial}{\partial\lambda}\ln a_1\right) \quad (112)$$

$$R^2{}_{303} = \frac{1}{2}[\ ]^{-1}\left(a_1\frac{\partial\omega_1}{c\partial\lambda}\right)\left(\frac{\partial}{\partial\lambda}\ln a_2\right) \quad (113)$$

$$R^2{}_{312} = -\frac{1}{2}[\ ]^{-1/2}\left(a_2\frac{\partial\omega_2}{c\partial\lambda}\right)\left(\frac{\partial}{c\partial t}\ln a_1\right) \quad (114)$$

$$R^2{}_{313} = \frac{1}{2}[\ ]^{-1/2}\left(a_1\frac{\partial\omega_1}{c\partial\lambda}\right)\left(\frac{\partial}{c\partial t}\ln a_2\right) \quad (115)$$

$$R^2{}_{323} = \left(\frac{\partial}{c\partial t}\ln a_1\right)\left(\frac{\partial}{c\partial t}\ln a_2\right) - [\ ]^{-1}\left(\frac{\partial}{\partial\lambda}\ln a_1\right)\left(\frac{\partial}{\partial\lambda}\ln a_2\right) \quad (116)$$

The remaining tetrad components of the Riemann tensor vanish. I turn thus to compute the Ricci tensor tetrad components from them by contractions on the pair of indices i and k. I find

$$R_{00} = -\frac{\partial^2}{c^2\partial t^2}\ln[\ ]^{1/2} - \left(\frac{\partial}{c\partial t}\ln[\ ]^{1/2}\right)^2 - \frac{\partial^2}{c^2\partial t^2}\ln a_1 - \left(\frac{\partial}{c\partial t}\ln a_1\right)^2 - \\ -\frac{\partial^2}{c^2\partial t^2}\ln a_2 - \left(\frac{\partial}{c\partial t}\ln a_2\right)^2 - \frac{1}{2}[\ ]^{-1}\left[\left(a_1\frac{\partial\omega_1}{c\partial\lambda}\right)^2 + \left(a_2\frac{\partial\omega_2}{c\partial\lambda}\right)^2\right] \tag{117}$$

$$R_{01} = [\ ]^{-1/2}\left\{\begin{array}{l}\left(\frac{\partial}{c\partial t}\ln[\ ]^{1/2}\right)\frac{\partial}{\partial\lambda}\ln a_1 + \left(\frac{\partial}{c\partial t}\ln[\ ]^{1/2}\right)\frac{\partial}{\partial\lambda}\ln a_2 - \\ -\frac{\partial^2}{\partial\lambda c\partial t}\ln a_1 - \left(\frac{\partial}{c\partial t}\ln a_1\right)\frac{\partial}{\partial\lambda}\ln a_1 - \frac{\partial^2}{\partial\lambda c\partial t}\ln a_2 - \left(\frac{\partial}{c\partial t}\ln a_2\right)\frac{\partial}{\partial\lambda}\ln a_2\end{array}\right\} \tag{118}$$

$$R_{02} = \frac{1}{2}[\ ]^{-1}\left\{\begin{array}{l}-\left(a_1\frac{\partial\omega_1}{c\partial\lambda}\right)\frac{\partial}{\partial\lambda}\ln[\ ]^{1/2} + \frac{\partial}{\partial\lambda}\left(a_1\frac{\partial\omega_1}{c\partial\lambda}\right) + \\ +2\left(a_1\frac{\partial\omega_1}{c\partial\lambda}\right)\frac{\partial}{\partial\lambda}\ln a_1\end{array}\right\} + \frac{1}{2}[\ ]^{-1}\left(a_1\frac{\partial\omega_1}{c\partial\lambda}\right)\frac{\partial}{\partial\lambda}\ln a_2 \tag{119}$$

$$R_{03} = \frac{1}{2}[\ ]^{-1}\left\{\begin{array}{l}-\left(a_2\frac{\partial\omega_2}{c\partial\lambda}\right)\frac{\partial}{\partial\lambda}\ln[\ ]^{1/2} + \frac{\partial}{\partial\lambda}\left(a_2\frac{\partial\omega_2}{c\partial\lambda}\right) + \\ +2\left(a_2\frac{\partial\omega_2}{c\partial\lambda}\right)\frac{\partial}{\partial\lambda}\ln a_2\end{array}\right\} + \frac{1}{2}[\ ]^{-1}\left(a_2\frac{\partial\omega_2}{c\partial\lambda}\right)\frac{\partial}{\partial\lambda}\ln a_1 \tag{120}$$

$$R_{11} = \frac{\partial^2}{c^2\partial t^2}\ln[\ ]^{1/2} + \left(\frac{\partial}{c\partial t}\ln[\ ]^{1/2}\right)^2 + \frac{1}{2}[\ ]^{-1}\left[\left(a_1\frac{\partial\omega_1}{c\partial\lambda}\right)^2 + \left(a_2\frac{\partial\omega_2}{c\partial\lambda}\right)^2\right] + \\ +\left(\frac{\partial}{c\partial t}\ln[\ ]^{1/2}\right)\frac{\partial}{c\partial t}\ln a_1 + \left(\frac{\partial}{c\partial t}\ln[\ ]^{1/2}\right)\frac{\partial}{c\partial t}\ln a_2 +$$

$$+[\ ]^{-1}\left\{\begin{bmatrix}\left(\frac{\partial}{\partial\lambda}\ln[\ ]^{1/2}\right)\frac{\partial}{\partial\lambda}\ln a_1-\frac{\partial^2}{\partial\lambda^2}\ln a_1-\left(\frac{\partial}{\partial\lambda}\ln a_1\right)^2+\\ +\left(\frac{\partial}{\partial\lambda}\ln[\ ]^{1/2}\right)\frac{\partial}{\partial\lambda}\ln a_2-\frac{\partial^2}{\partial\lambda^2}\ln a_2-\left(\frac{\partial}{\partial\lambda}\ln a_2\right)^2\end{bmatrix}\right\} \qquad (121)$$

$$R_{12}=\frac{1}{2}[\ ]^{-1/2}\left\{\begin{bmatrix}-\left(a_1\frac{\partial\omega_1}{c\partial\lambda}\right)\frac{\partial}{c\partial t}\ln[\ ]^{1/2}+\frac{\partial}{c\partial t}\left(a_1\frac{\partial\omega_1}{c\partial\lambda}\right)+\\ +2\left(a_1\frac{\partial\omega_1}{c\partial\lambda}\right)\frac{\partial}{c\partial t}\ln a_1\end{bmatrix}\right\}+\frac{1}{2}[\ ]^{-1/2}\left(a_1\frac{\partial\omega_1}{c\partial\lambda}\right)\frac{\partial}{c\partial t}\ln a_2 \qquad (122)$$

$$R_{13}=\frac{1}{2}[\ ]^{-1/2}\left\{\begin{bmatrix}-\left(a_2\frac{\partial\omega_2}{c\partial\lambda}\right)\frac{\partial}{c\partial t}\ln[\ ]^{1/2}+\frac{\partial}{c\partial t}\left(a_2\frac{\partial\omega_2}{c\partial\lambda}\right)+\\ +2\left(a_2\frac{\partial\omega_2}{c\partial\lambda}\right)\frac{\partial}{c\partial t}\ln a_2\end{bmatrix}\right\}+\frac{1}{2}[\ ]^{-1/2}\left(a_2\frac{\partial\omega_2}{c\partial\lambda}\right)\frac{\partial}{c\partial t}\ln a_1 \qquad (123)$$

$$R_{22}=\frac{\partial^2}{c^2\partial t^2}\ln a_1+\left(\frac{\partial}{c\partial t}\ln a_1\right)^2+[\ ]^{-1}\left\{\left(\frac{\partial}{\partial\lambda}\ln[\ ]^{1/2}\right)\frac{\partial}{\partial\lambda}\ln a_1-\frac{\partial^2}{\partial\lambda^2}\ln a_1-\left(\frac{\partial}{\partial\lambda}\ln a_1\right)^2\right\}+$$

$$+\left(\frac{\partial}{c\partial t}\ln[\ ]^{1/2}\right)\frac{\partial}{c\partial t}\ln a_1+\left(\frac{\partial}{c\partial t}\ln a_1\right)\left(\frac{\partial}{c\partial t}\ln a_2\right)-[\ ]^{-1}\left(\frac{\partial}{\partial\lambda}\ln a_1\right)\left(\frac{\partial}{\partial\lambda}\ln a_2\right)-\frac{1}{2}[\ ]^{-1}\left(a_1\frac{\partial\omega_1}{c\partial\lambda}\right)^2 \qquad (124)$$

$$R_{23}=-\frac{1}{2}[\ ]^{-1}\left(a_1\frac{\partial\omega_1}{c\partial\lambda}\right)\left(a_2\frac{\partial\omega_2}{c\partial\lambda}\right) \qquad (125)$$

$$R_{33}=\frac{\partial^2}{c^2\partial t^2}\ln a_2+\left(\frac{\partial}{c\partial t}\ln a_2\right)^2+[\ ]^{-1}\left\{\left(\frac{\partial}{\partial\lambda}\ln[\ ]^{1/2}\right)\frac{\partial}{\partial\lambda}\ln a_2-\frac{\partial^2}{\partial\lambda^2}\ln a_2-\left(\frac{\partial}{\partial\lambda}\ln a_2\right)^2\right\}+$$

$$+\left(\frac{\partial}{c\partial t}\ln[\ ]^{1/2}\right)\frac{\partial}{c\partial t}\ln a_2+\left(\frac{\partial}{c\partial t}\ln a_1\right)\left(\frac{\partial}{c\partial t}\ln a_2\right)-[\ ]^{-1}\left(\frac{\partial}{\partial\lambda}\ln a_1\right)\left(\frac{\partial}{\partial\lambda}\ln a_2\right)-\frac{1}{2}[\ ]^{-1}\left(a_2\frac{\partial\omega_2}{c\partial\lambda}\right)^2. \qquad (126)$$

Finally, contracting the Ricci tensor, I find for the scalar curvature

$$R = -2\frac{\partial^2}{c^2\partial t^2}\ln[\;]^{1/2} - 2\left(\frac{\partial}{c\partial t}\ln[\;]^{1/2}\right)^2 - 2\frac{\partial^2}{c^2\partial t^2}\ln a_1 - 2\left(\frac{\partial}{c\partial t}\ln a_1\right)^2 -$$

$$-2\frac{\partial^2}{c^2\partial t^2}\ln a_2 - 2\left(\frac{\partial}{c\partial t}\ln a_2\right)^2 - \frac{1}{2}[\;]^{-1}\left[\left(a_1\frac{\partial\omega_1}{c\partial\lambda}\right)^2 + \left(a_2\frac{\partial\omega_2}{c\partial\lambda}\right)^2\right] -$$

$$-2\left(\frac{\partial}{c\partial t}\ln[\;]^{1/2}\right)\frac{\partial}{c\partial t}\ln a_1 - 2\left(\frac{\partial}{c\partial t}\ln[\;]^{1/2}\right)\frac{\partial}{c\partial t}\ln a_2 -$$

$$-2\left(\frac{\partial}{c\partial t}\ln a_1\right)\left(\frac{\partial}{c\partial t}\ln a_2\right) - [\;]^{-1}\left\{\begin{array}{l} 2\left(\frac{\partial}{\partial\lambda}\ln[\;]^{1/2}\right)\frac{\partial}{\partial\lambda}\ln a_1 - 2\frac{\partial^2}{\partial\lambda^2}\ln a_1 - 2\left(\frac{\partial}{\partial\lambda}\ln a_1\right)^2 + \\ + 2\left(\frac{\partial}{\partial\lambda}\ln[\;]^{1/2}\right)\frac{\partial}{\partial\lambda}\ln a_2 - 2\frac{\partial^2}{\partial\lambda^2}\ln a_2 - 2\left(\frac{\partial}{\partial\lambda}\ln a_2\right)^2 - \\ -2\left(\frac{\partial}{\partial\lambda}\ln a_1\right)\left(\frac{\partial}{\partial\lambda}\ln a_2\right) \end{array}\right\}. \quad (128)$$

**Appendix IV:** *The curvature tensor via the λ-symbols of tetrad formalism*

The λ-symbols of tetrad formalism (Landau & Lifshitz, 1975; Chandrasekhar, 1992) can be defined as

$$\lambda_{(a)(b)(c)} = e_{(a)}{}^{i}\left[e_{(b)i,j} - e_{(b)j,i}\right]e_{(c)}{}^{j}, \quad (129)$$

where the comma (,) denotes partial (<u>not</u> covariant) differentiation. It is evident that

$$\lambda_{(a)(b)(c)} = -\lambda_{(c)(b)(a)}. \quad (130)$$

The tetrad components of the Ricci tensor can then be expressed through the λ-symbols (Landau & Lifshitz, 1975) as

$$R_{(a)(b)} = -\frac{1}{2}\left(\begin{array}{l} \lambda_{(a)(b)}{}^{(c)}{}_{,(c)} + \lambda_{(b)(a)}{}^{(c)}{}_{,(c)} + \lambda_{(c)}{}^{(c)}{}_{(a),(b)} + \lambda_{(c)}{}^{(c)}{}_{(b),(a)} + \lambda^{(c)(d)}{}_{(b)}\lambda_{(c)(d)(a)} + \\ + \lambda^{(c)(d)}{}_{(b)}\lambda_{(d)(c)(a)} - \frac{1}{2}\lambda^{(c)}{}_{(b)}{}^{(d)}\lambda_{(c)(a)(d)} + \lambda_{(c)}{}^{(c)}{}_{(d)}\lambda_{(a)(b)}{}^{(d)} + \lambda_{(c)}{}^{(c)}{}_{(d)}\lambda_{(b)(a)}{}^{(d)} \end{array}\right), \quad (131)$$

where the raising (and lowering) of tetrad indices is performed with $\eta^{ab}$ (and $\eta_{ab}$), and the comma (,) here denotes <u>directional</u> derivation.

Since we have a complete knowledge of the tetrad basis (see subsection iii)), all we need is finding of the partial derivatives of these basis vectors, appearing inside the square bracket in eqn. (129). I find, for the non-vanishing of them, for b = 1

$$e_{(1)1,0} = -\frac{\partial}{c\partial t}[\ ]^{1/2}$$
$$e_{(1)1,1} = -\frac{\partial}{\partial\lambda}[\ ]^{1/2}, \tag{132}$$

for b = 2

$$e_{(2)0,0} = \frac{\partial a_1}{c\partial t}\frac{\omega_1}{c} + a_1\frac{\partial\omega_1}{c^2\partial t}$$
$$e_{(2)0,1} = \frac{\partial a_1}{\partial\lambda}\frac{\omega_1}{c} + \left(a_1\frac{\partial\omega_1}{c\partial\lambda}\right) \tag{133}$$

and

$$e_{(2)2,0} = -\frac{\partial a_1}{c\partial t}$$
$$e_{(2)2,1} = -\frac{\partial a_1}{\partial\lambda}, \tag{134}$$

and for b = 3

$$e_{(3)0,0} = \frac{\partial a_2}{c\partial t}\frac{\omega_2}{c} + a_2\frac{\partial\omega_2}{c^2\partial t}$$
$$e_{(3)0,1} = \frac{\partial a_2}{\partial\lambda}\frac{\omega_2}{c} + \left(a_2\frac{\partial\omega_2}{c\partial\lambda}\right) \tag{135}$$

and

$$e_{(3)3,0} = -\frac{\partial a_2}{c\partial t}$$
$$e_{(3)3,1} = -\frac{\partial a_2}{\partial\lambda} \tag{136}$$

(for b = 0 non of the partial derivatives is non-vanishing).

For the square bracket in eqn. (129) (if it is non-vanishing) I find, for b = 1

$$e_{(1)0,1} - e_{(1)1,0} = \frac{\partial}{c\partial t}[\ ]^{1/2} = -[e_{(1)1,0} - e_{(1)0,1}], \tag{137}$$

for b = 2

$$e_{(2)0,1} - e_{(2)1,0} = \frac{\partial a_1}{\partial \lambda}\frac{\omega_1}{c} + \left( a_1 \frac{\partial \omega_1}{c\partial \lambda} \right) = -\left[ e_{(2)1,0} - e_{(2)0,1} \right] \tag{138}$$

$$e_{(2)0,2} - e_{(2)2,0} = \frac{\partial a_1}{c\partial t} = -\left[ e_{(2)2,0} - e_{(2)0,2} \right] \tag{139}$$

$$e_{(2)1,2} - e_{(2)2,1} = \frac{\partial a_1}{\partial \lambda} = -\left[ e_{(2)2,1} - e_{(2)1,2} \right], \tag{140}$$

and for b = 3

$$e_{(3)0,1} - e_{(3)1,0} = \frac{\partial a_2}{\partial \lambda}\frac{\omega_2}{c} + \left( a_2 \frac{\partial \omega_2}{c\partial \lambda} \right) = -\left[ e_{(3)1,0} - e_{(3)0,1} \right] \tag{141}$$

$$e_{(3)0,3} - e_{(3)3,0} = \frac{\partial a_2}{c\partial t} = -\left[ e_{(3)3,0} - e_{(3)0,3} \right] \tag{142}$$

$$e_{(3)1,3} - e_{(3)3,1} = \frac{\partial a_2}{\partial \lambda} = -\left[ e_{(3)3,1} - e_{(3)1,3} \right] \tag{143}$$

(again for b = 0 the result is null).

Then I compute the λ-symbols from eqn. (129). The non-vanishing of them are:

for b = 1

$$\lambda_{(0)(1)(1)} = \frac{\partial}{c\partial t}\ln[\ ]^{1/2} = -\lambda_{(1)(1)(0)}, \tag{144}$$

for b = 2

$$\lambda_{(0)(2)(1)} = [\ ]^{-1/2}\left( a_1 \frac{\partial \omega_1}{c\partial \lambda} \right) = -\lambda_{(1)(2)(0)} \tag{145}$$

$$\lambda_{(0)(2)(2)} = \frac{\partial}{c\partial t}\ln a_1 = -\lambda_{(2)(2)(0)} \tag{146}$$

$$\lambda_{(1)(2)(2)} = [\ ]^{-1/2}\frac{\partial}{\partial \lambda}\ln a_1 = -\lambda_{(2)(2)(1)}, \tag{147}$$

and for b = 3

$$\lambda_{(0)(3)(1)} = [\ ]^{-1/2}\left( a_2 \frac{\partial \omega_2}{c\partial \lambda} \right) = -\lambda_{(1)(3)(0)} \tag{148}$$

$$\lambda_{(0)(3)(3)} = \frac{\partial}{c\partial t}\ln a_2 = -\lambda_{(3)(3)(0)} \tag{149}$$

$$\lambda_{(1)(3)(3)} = [\ ]^{-1/2} \frac{\partial}{\partial\lambda} \ln a_2 = -\lambda_{(3)(3)(1)} \tag{150}$$

Then, from the λ-symbols found, I easily compute the various combinations of them appearing in eqn. (131). Note that

$$\lambda_{(a)(b)}{}^{(c)}{}_{,(c)} = \lambda_{(a)(b)}{}^{(c)}{}_{,i} e_{(c)}{}^{i} \tag{151}$$

and

$$\lambda_{(c)}{}^{(c)}{}_{(a),(b)} = \lambda_{(c)}{}^{(c)}{}_{(a),i} e_{(b)}{}^{i}. \tag{152}$$

Finally, from formula (131), I compute the tetrad components of the Ricci tensor themselves. I do not list them, since they agree with the ones found in the previous subsection (*iv)*). And similarly for the scalar curvature R I find of course the same result with the one previously found, given by eqn. (128).

**Appendix V:** *The energy-momentum tensor and Einstein´s equations*

I assume the cosmic fluid to be a comoving (with the coordinates) perfect fluid. In this case its energy-momentum tensor will be given by

$$T^{ik} = (p+\varepsilon)u^i u^k - p g^{ik}, \tag{153}$$

where ε is the fluid´s energy density and p its pressure, with the four-velocity $u^i$, defined by

$$u^i = \frac{dx^i}{ds} \tag{154}$$

being given by

$$u^\alpha = 0, \qquad u^0 = 1/\sqrt{g_{00}} \tag{155}$$

(the latin indices refer to space-time, taking the values 0, 1, 2, 3, while the greek indices refer only to spatial coordinates, taking the values 1, 2, 3, as always). Of course $g_{00}$ is given by

$$g_{00} = 1 - a_1^2 \frac{\omega_1^2}{c^2} - a_2^2 \frac{\omega_2^2}{c^2} \tag{156}$$

(cf. the first of eqns. (41)). Then the tensor components of $u^i$ will be

$$u^i = \left(1/\sqrt{g_{00}}, 0, 0, 0\right). \tag{157}$$

But I am interested in the *tetrad* components $u^{(a)}$ of the four-velocity. These are given by

$$u^{(a)} = e^{(a)}{}_i u^i = e^{(a)}{}_0 / \sqrt{g_{00}}\,. \tag{158}$$

And, since

$$e^{(a)}{}_0 = \left(1, 0, -a_1 \omega_1/c, -a_2 \omega_2/c\right) \tag{159}$$

(cf. eqns. (49)), I will have

$$u^{(a)} = \left(\frac{1}{\sqrt{g_{00}}}, 0, \frac{-a_1 \omega_1/c}{\sqrt{g_{00}}}, \frac{-a_2 \omega_2/c}{\sqrt{g_{00}}}\right). \tag{160}$$

Thus the *tetrad* components of the energy-momentum tensor will be given by

$$T^{(a)(b)} = (p+\varepsilon) u^{(a)} u^{(b)} - p\eta^{ab}, \tag{161}$$

with the four-velocity (*tetrad* components) given by eqn. (160) and $\eta^{ab}$ by

$$\eta^{ab} = diag(+1, -1, -1, -1), \tag{162}$$

that is Minkowskian (cf. eqn. (54)).

I get

$$T^{(0)(0)} = (p+\varepsilon)\frac{1}{g_{00}} - p$$

$$T^{(0)(1)} = T^{(1)(0)} = 0$$

$$T^{(0)(2)} = T^{(2)(0)} = -\frac{a_1\omega_1/c}{g_{00}}(p+\varepsilon)$$

$$T^{(0)(3)} = T^{(3)(0)} = -\frac{a_2\omega_2/c}{g_{00}}(p+\varepsilon)$$

$$T^{(1)(1)} = p$$

$$T^{(1)(2)} = T^{(2)(1)} = 0$$

$$T^{(1)(3)} = T^{(3)(1)} = 0$$

$$T^{(2)(2)} = \frac{a_1^{\;2}\omega_1^{\;2}/c^2}{g_{00}}(p+\varepsilon) + p$$

$$T^{(2)(3)} = T^{(3)(2)} = \frac{a_1 a_2\omega_1\omega_2/c^2}{g_{00}}(p+\varepsilon)$$

$$T^{(3)(3)} = \frac{a_2^{\;2}\omega_2^{\;2}/c^2}{g_{00}}(p+\varepsilon) + p. \tag{163}$$

For the scalar T, given by

$$T = T^{(a)}{}_{(a)} = \eta_{ab}T^{(a)(b)}, \tag{164}$$

I obtain

$$T = \varepsilon - 3p. \tag{165}$$

After those, the Einstein equations in tetrad form can be written in the form (Landau & Lifshitz, 1975)

$$R^{(a)(b)} = \frac{8\pi G}{c^4}\left(T^{(a)(b)} - \frac{1}{2}\eta^{ab}T\right) \tag{166}$$

$$T^{(0)(0)} - \frac{1}{2}\eta^{00}T = \frac{1}{g_{00}}(p+\varepsilon) + \frac{1}{2}(p-\varepsilon)$$

$$T^{(0)(1)} - \frac{1}{2}\eta^{01}T = 0$$

$$T^{(0)(2)} - \frac{1}{2}\eta^{02}T = -\frac{a_1\,\omega_1/c}{g_{00}}(p+\varepsilon)$$

$$T^{(0)(3)} - \frac{1}{2}\eta^{03}T = -\frac{a_2\,\omega_2/c}{g_{00}}(p+\varepsilon)$$

$$T^{(1)(1)} - \frac{1}{2}\eta^{11}T = -\frac{1}{2}(p-\varepsilon)$$

$$T^{(1)(2)} - \frac{1}{2}\eta^{12}T = 0$$

$$T^{(1)(3)} - \frac{1}{2}\eta^{13}T = 0$$

$$T^{(2)(2)} - \frac{1}{2}\eta^{22}T = \frac{a_1^{\;2}\,\omega_1^{\;2}/c^2}{g_{00}} - \frac{1}{2}(p-\varepsilon)$$

$$T^{(2)(3)} - \frac{1}{2}\eta^{23}T = \frac{a_1 a_2\,\omega_1\omega_2/c^2}{g_{00}}(p+\varepsilon)$$

$$T^{(3)(3)} - \frac{1}{2}\eta^{33}T = \frac{a_2^{\;2}\,\omega_2^{\;2}/c^2}{g_{00}}(p+\varepsilon) - \frac{1}{2}(p-\varepsilon). \qquad (167)$$

I have altogether ten (10) Einstein equations. The unknowns determining the metric are the four (4) functions $a_1(\lambda, t)$, $a_2(\lambda, t)$; $\omega_1(\lambda, t)$, $\omega_2(\lambda, t)$. But, as it will be seen in the sequel, these functions can be separated in λ & t. Thus, I will have 4×2 = 8 unknown functions determining the metric.

Concerning the four-velocity, I must have in principle 3 unknowns determing it (for example its spatial components), left after using the identity

$$u^{(a)}u_{(a)} = 1. \qquad (168)$$

But I have forced the spatial components to be zero *ad hoc,* without destroying the equality (168) as it can be easily seen. Thus, the four-velocity contains no additional unknowns.

There are, nevertheless, two additional unknowns in the energy-momentum tensor: $\varepsilon$ & p. Normally only one of them can be considered as an unknown, the other one resulting by the equation of state (an extra equation used). Thus one Einstein equation remains, which has to be satisfied identically. But we can use it in order to *determine* the equation of state, instead of imposing the latter. Thus we see that here the equation of state is *contained* in the field equations!

I should list the Einstein equations. But this is forbidable because of the limiting space available. Thus the reader has to form them by himself, by equating the various tetrad components of the Ricci tensor exposed in subsection (iv) to the corresponding right-hand sides listed in the list of eqns. (167), after raising or lowering indices with the help of the “metric” $\eta^{ab}$ or $\eta_{ab}$.

# REFERENCES

## A. Books

1. Binney, J. & Tremaine, S. 1987, "Galactic Dynamics", Princeton, Princeton
2. Chandrasekhar, S. 1992, "The Mathematical Theory of Black Holes", Clarendon, Oxford
3. Chevalley, C. 1946, "Theory of Lie Groups", Princeton, Princeton
4. Contopoulos, G. & Kotsakis, D. 1987, « Cosmology », Springer-Verlag, Berlin
5. Hawking, S.W. & Ellis, G.F.R. 1973, "The Large Scale Structure of Space-Time", Cambridge, Cambridge
6. Kappos, D. 1966, "Differential Equations", Athens (Greece) (in Greek)
7. Kramer, D., Stephani, H.,Hertl, E., MacCallum, M., Schmutzer, E. 1980, "Exact Solutions of Einstein´s Field Equations", Cambridge, Cambridge
8. Landau, L.D. & Lifshitz, E.M. 1975, "The Classical Theory of Fields", 4th edn., vol. 2 of "Course of Theoretical Physics", Pergamon, Oxford
9. Lovelock, D. & Rund, H. 1975, "Tensors, Differential Forms, and Variational Principles", Wiley, New York
10. Ryan, M.P., Jr. & Shepley, L.C. 1975, "Homogeneous Relativistic Cosmologies", Princeton, Princeton

## B. Papers

1. Carroll, S.M., Press, W.H. & Turner, E.L. 1992, *ARA&A,* **30,** 499
2. Chaliasos, E., 2006a; http://arxiv.org/abs/physics/0607214
3. Chaliasos, E., 2006b; http://arxiv.org/abs/physics/0609122
4. Glanz, J., 1998; *Science* **279** 1298
5. Koppar, S.S. & Patel, L.K., 1988a; *Nuovo Cimento* **102** 419
6. Koppar; S.S. & Patel, L.K., 1988b; *Nuovo Cimento* **102** 425
7. Korotkii, V.A. & Obukhov, Yu.N., 1992; *Astrophys. Sp. Sc.* **198** 1
8. Novello, M. & Reboucas, M.J., 1978; *Ap.J.* **225** 719
9. Novello, M. & Reboucas, M.J., 1979; *Phys.Rev.D* **19** 2850
10. Perlmutter, S., *et al,* 1998 ; *Nature* **391** 51
11. Reboucas, M.J., 1979 ; *Phys.Lett.* **70A** 161
12. Reboucas, M.J., 1982 ; *Nuovo Cimento* **67B** 120
13. Reboucas, M.J. & de Lima, J.A.S., 1981; *J.Math.Phys.* **22**(11) 2699
14. Reboucas, M.J. & Tiomno, J., 1983; *Phys.Rev.D* **28** 1251
15. Rosquist, K., 1983 ; *Phys.Lett.* **97A** 145
16. Usham, J.K. & Tarachand Singh, R.K., 1992; *Astrophys.Sp.Sc.* **189** 79
17. Vaidya, P.C. & Patel, L.K., 1984; *Gen.Rel.Grav.* **16** 355